\documentclass{article}

\usepackage{arxiv}

\usepackage[utf8]{inputenc} 
\usepackage[T1]{fontenc}    
\usepackage{hyperref}       
\usepackage{url}            
\usepackage{booktabs}       
\usepackage{amsfonts}       
\usepackage{nicefrac}       
\usepackage{microtype}      
\usepackage{lipsum}
\usepackage{graphicx}
\usepackage{authblk}

\title{How Segregation Patterns Affect the Availability of Fair District Plans}

\author{William Hager \qquad Betseygail Rand \\  Dept. of Mathematics, Computer Science, and Information Systems, \\ Texas Lutheran University, 1000 W Court St., Seguin, 78155, Texas, USA}

\begin{document}

\maketitle

\begin{abstract}
   We create 4200 synthetic cities which vary in percent minority population and their residential segregation patterns.  Of these, 1200 are modeled on existing cities, and 3000 are rectangular grid cities. In each city, we consider single-member voting district plans for a hypothetical city council election.  A fair district plan is defined as one where the number of minority-majority districts is proportional to the city-wide minority population. Thus each city is summarized by three traits: minority percent, a measure of segregation, and availability of a fair district plan. We find that when the minority population is around 25\%-33\%, there is a positive correlation between the degree of segregation and the availability of proportional district plan.  Consistently, when the minority population lives in a more diffuse residential pattern, there are fewer available proportional district plans. Finally, we develop a new method to validate runtime and sample size of an ensemble of district plans created by the GerryChain software program. 
\end{abstract}


\section{Introduction}

In this paper, we investigate the question of how residential segregation patterns affect the availability of a fair district plan. We are interested in city council districts of midsize American cities, roughly 300,000-750,000 in population, with a city council elected by single member districts, ranging from 7-13 districts in the city. We create 4200 synthetic cities, some modeled on specific American cities and others based on a rectangular grid. Each city has three associated properties: the percent minority of total population, a measure of racial segregation, and a measure of the availability of a proportional district plan. 

We find a positive correlation between the degree of residential segregation and the availability of a fair district plan. Moreover, a large proportion of cities have no proportional plans available at all. We conclude that in a majority of cases, single member districts are likely not the best method to ensure proportional representation.\footnote{The authors wish to acknowledge a debt to Moon Duchin, whose offhand comment during the Austin Gerrymandering Conference (February 3rd, 2018) planted the question for study.}

Along the way, we propose a new method of validating sample size for an ensemble analysis of valid district plans.  We create a definition for two district plans to be {\em distant} from each other, and find an upper bound for a maximal set of mutually distant plans.  We then locate a maximal number of mutually distant plans in the ensemble plans.  We find that our ensembles achieve between 0.53 and 0.85 of the possible upper bounds, providing some evidence that the statespace is in fact being explored. Implementing this validation requires very little computational power.

In Section \ref{measures of gerrymandering}, we first review existing measures of quantifying gerrymandering. We then define $\bar F$, the fairness index, a measure of the availability of a proportional district plan, using an ensemble analysis. In Section \ref{measures of segregation}, we review measures of segregation in the literature, and introduce the choice for this paper, $D$, the dissimilarity index. In Section \ref{methodology}, we describe our methodology of creating the synthetic cities. Of these, 3000 are rectangular grid cities, and 1200 are modeled after American cities. We also give a new method for validating runtime of GerryChain in this section. In Section \ref{results}, we give the results and establish strong correlations between the degree of segregation and the availability of a proportional district plan. In Section \ref{context}, we ground these results in the broader current American context.

\section{Measures of Gerrymandering}\label{measures of gerrymandering}

It is useful to separate out different categories used when evaluating gerrymandering. First, people differ on what is meant by a fair district plan. Second, there is a choice on how to evaluate fairness. Third, there is a separate question of how to assess the availability of a fair district plan for a specific election.

What is a reasonable notion of {\em fair}?  First, each single-member district must have equal population, under the Equal Protection Clause, and must comply with the Voting Rights Act.   Beyond that, there are many different metrics to evaluate a district plan for gerrymandering, and often these have a specific meaning of ``fair" embedded within them. Historically, ``compactness" was taken as a measure of fairness, and oddly-shaped districts were suspected of being gerrymandered. This speaks to a meaning of ``fair" based on human behavior: small-scale manipulation of boundaries is unfair and indicates an attempt to swing an election.  More recently, there has been an effort to quantify ``fairness" more precisely. For example, the efficiency gap measures the excess votes of the winning party against the votes that did not contribute to a win of the losing party. \cite{duchin2018gerrymandering,tapp2019measuring} Underlying this, ``fair" implies that the two parties ought to have the same proportion  of voters whose votes contribute to victories.  For measures based on a seats-vote curve, ``fair" is a linear vote distribution. \cite{warrington2019comparison} Mean-median bias, partisan bias, and other symmetry based measures take into account the deviation of an election outcome from this linear vote distribution.

Partisan gerrymandering is nonjusticiable under current law, \cite{RC19} but could be considered easier to detect than racial gerrymandering, because detecting party preferences is reasonably straightforward - the outcome of an election is often a suitable proxy for partisan allegiance. The outcome of a single election can be taken as a basis for a seats-votes curve or to measure an efficiency gap. Racial gerrymandering is legally protected, but harder to detect.  Under the 1986 decision in Thornberg vs. Gingles, there were established some preconditions for racial gerrymandering, namely the degree of racially polarized voting in a region and voter turnout by racial subgroup. \cite{Bpedia}  The outcome of a single election does not give enough information to determine if the majority subgroup is consistently preventing a minority subgroup from being represented by a candidate of choice.

In this paper, we are interested in ``fair" as it applies to racial bias in district plans. We will not attempt to model racially polarized voting, nor varying degrees of voter turnout. We will also not model elections.  Instead our notion of ``fair" is based on counting minority-majority districts within a district plan. A fair district plan will be defined as one where the number of minority-majority districts is proportional to the minority population of the city. We will use census data, rather than Voting Age Population (VAP) data. This is not identical to measuring opportunity districts, which are determined using voter turnout and election outcomes, but our methods could be adjusted for a specific city where the threshold for being an opportunity district has been established.  One additional advantage is that determining a minority-majority district is essentially the same computation as a single-member district election outcome, in the context of studying partisan gerrymandering. In the next section, we formalize our notion of ``fair".

\subsection{The Fairness Index, $\bar F$}

Suppose we have a city $X$ with $n$ districts, with population $P$.  We designate the historically  vulnerable subgroup $Q$,  and we require $Q > P/2n$, the minimum sufficient number of votes to constitute a majority of one district.  We consider $P-Q$ to be the remainder group, including both the traditionally dominant population and all other minority populations.  As a simplification, we will almost entirely restrict our attention to cities with only one historically vulnerable subgroup that is sufficiently large to win a district. For a given district plan $d$, let $d_Q$ be the number of districts which are minority-majority.

To determine if this percent is fair, we appeal to our intuition: a district plan is {\em fair} if the percent minority-majority districts matches the city-wide minority $Q/P$.  For example, suppose a city is 30\% African-American and has 10 single-member voting districts. Under district plan $d$, there are three minority-majority districts, and so $d_Q=3$.  This achieves proportional representation, and thus is fair.  In contrast, suppose a second district plan $d'$ has one minority-majority district, $d'_Q=1$. This falls short of proportional representation, and thus is not fair. 

We define the fairness index $F(d,Q,X) = \frac{d_Q/n}{Q/P}$, the ratio of the percent minority-majority districts to the percent population $Q$. For the example above, $F(d,Q,X) = 1$ and $F(d',Q,X)=1/3$.
In general, if $F < 1$, then population $Q$ is underrepresented in the district plan $d$, and if $F > 1$, then population $Q$ is overrepresented in minority-majority districts in $d$. 

We note that $F(d,Q,X)$ is useful for evaluating representation of a subgroup $Q$ across different district plans and across different cities, but it is not symmetric about 1 in an immediately useful way. If $Q/P = 0.3$ and $d_Q/n = 0.4$, then  $F(d,Q,X) = 1.33$, while $F(d,P-Q,X) = 0.85$.

\subsubsection{$\bar F$, the city-wide Fairness Index} 

The question arises: is a fair district plan always available? Is it sometimes the case that a fair district plan cannot be drawn? This question has been well-answered by the development of ensemble analyses, which determine a baseline availability of a district plan according to a given metric of fairness. \cite{duchin2018gerrymandering}

We use GerryChain, the software package developed by the Metric Geometry and Gerrymandering Group (MGGG), in 2017.\cite{GerryChain} For each city, the program generates 20,000 valid district plans. (For a district plan to be valid, the districts must all be continuous, and within 0.2 of the average district population.) See Appendix \ref{gerrychain settings} for details.

For a fixed city $X$, and a set of district plans $S$, $F_d$ is computed for all district plans $d \in S$, with respect to a subgroup $Q$. We then define $\bar F$ for this city to be the average of  $\{ F_d \vert d \in S \}$,ie $\bar F(X, S, Q) =  \sum_{d\in S} \frac{F_d}{\vert S \vert}$.  If $X$, $S$  and $Q$ are unambiguous from context, we will use $\bar F$ without indicating these.

If $\bar F =1 $, then we use this as a proxy to presume that a map yielding a proportional number of minority-majority districts is readily available.  If $\bar F < 1$, then it will be more difficult to locate a fair map. If $\bar F > 1$, then the subgroup $Q$ is overrepresented in the majority of district plans.  

The program must run long enough to ensure that the random walk thoroughly explores the statespace, and generates a statistically representative sample. In Section \ref{validate}, we develop a new method to validate the runtime of GerryChain. Briefly, the statespace of district plans is impossibly large \cite{MGGG2}, and we propose a definition for two district plans to be {\em distant} in this statespace. We then find an upper bound on the size of a set of pairwise distant district plans, and then determine the number of pairwise distant district plans in the set of 20,000 plans.  We find that our set of district plans has a number of distant district plans which is close to maximal, indicating that the set is in fact sampling many parts of the statespace.

\section{Measures of Segregation} \label{measures of segregation}

\subsection{Brief Literature Review}

There is extensive literature on quantitative measures of segregation. One of the oldest measures of segregation is the dissimilarity index, proposed by Duncan \& Duncan in 1955. \cite{DD55} This measure computes the region-wide proportion of two groups of people, and then aggregates the local deviation of proportion, across small spatial units that make up the region. To date, this is the most commonly used measure of segregation. In a very influential 1988 paper, Massey \& Denton described five qualities that constitute racial segregation: evenness, exposure, concentration, centralization, and clustering, and surveyed 20 measures of segregation according to their performance on these measures.  \cite{MD88} The dissimilarity index is considered to measure evenness. They conclude that $D$ correlates with measures of the other qualities, and thus affirm the practice of using it as a summary measure, especially given how simple a computation it is.  However, $D$ does not extend easily to multi-group situations. As an indication of its influence, the Census Bureau organizes its webpage on Measures of Residential Segregation according to Massey \& Denton, and gives formulas as presented in their 1988 paper. \cite{USCB1}

Two major critiques of traditional measures are known as the Checkerboard Problem and the Modifiable Unit Area Problem. The Checkerboard Problem refers to the fact that $D$ is the sum of computations which are each contained entirely within their own areal unit. Adjacencies of areal units do not contribute to the computation. \cite{W83} Therefore, any rearrangement of areal units will produce a region with the same value of $D$.  In particular, all similar units could be clustered together in an extreme fashion (for example, splitting the board half black and half white) or distributed evenly, like a checkerboard, and the value of $D$ would not vary. The Modifiable Area Unit Problem (MAUP) was first described by Gehlke and Beale \cite{GB34} and describes the general statistical problem of how aggregating geographic data can be affected by choice of scale and boundary. Geographic indices like $D$ depend on the choice of boundaries when aggregating data, and redrawing the boundaries that are used to aggregate residential statistics can produce varying values for $D$ and other measures. For example, suppose the actual living patterns of residents resemble a checkerboard. If census tracts are drawn to line up with the underlying checkerboard, then all areal units will contribute maximal deviation from the citywide average. If census tracts are drawn to straddle adjacent squares, then each areal unit would not deviate at all from the citywide average, and would contribute 0.  In practice, a refinement of areal unit will lead to larger segregation index values, \cite{YWBM19}.

To address these concerns, spatial measures of segregation have been developed \cite{YWBM19}. Many of these modify their metric by allowing the local areal computation to incorporate the population ratios the neighboring areal units, which is then weighted according to the population of the small unit at the center of the region.  In this way, a region contributes to all its neighbors. However, no single measure newer has emerged as a widespread standard, and all of them require a substantial increase in computations over traditional measures.

\subsection{Choice of Segregation Measure}

We use $D$, the dissimilarity index, to measure the segregation of our synthetic cities. In general, suppose $Q$ and $P-Q$ are mutually exclusive populations in a city $X$.  Suppose further that the city is partitioned into areal tracts $\{v_i\}$, and $Q_i$ and $(P-Q)_i$ are the respective populations of $Q$ and $P-Q$ living within $v_i$. The dissimilarity index $D$ is computed by:

$$D = \frac{1}{2}\sum_i\left\vert \frac{Q_i}{Q}-\frac{(P-Q)_i}{P-Q}\right\vert$$

\cite{USCB1}.
 If the ratio of $Q_i$ and $(P-Q)_i$ living in tract $v_i$ match the citywide ratio of $Q$ and $P-Q$, then that areal tract $v_i$ contributes 0 to the total $D$. If the tract $v_i$ is composed homogenously of one group, and has no members of the other living there, then that tract will contribute maximally to the sum $D$. Therefore, $D$ measures the extent to which areal tracts deviate from the city-wide proportion of races. A city in which every tract is homogeneous will have $D =1$, and a city where every tract is proportional to the city will have $D=0$.

\section{Methodology} \label{methodology}

\subsection{Building Synthetic Cities}

In order to explore the relationship between racial residential patterns and the availability of fair district plans, we create 4200  cities with varying degrees of segregation, and compute the fairness index for each. Of these, 1200 cities are intended to be realistic and complex, and are modeled after real cities. The remaining 3000 are simpler and more artificial - these are rectangular grid cities, with the same fixed population at each census block.  

\subsubsection{Synthetic Modeled Cities}

In designing realistic cities, we confront infinitely many parameters that may describe a city. What does it mean to have a realistic boundary or shape to a city? Should geographic features like rivers or mountains be present? Where should an urban core be located? Where should the populations be densest and least dense? What should the ratio of $Q$ to $P$ be, city-wide? How  should the ratio of $Q$ to $P$ vary at each census tract, in order to vary racial residential patterns? How many single-member districts should the city be divided into, or equivalently, how many individuals should be in a district? What is the number of votes needed to win a single district, and how does that compare to the size of the minority population? 

To simplify the enormous complexity of this question, we choose four cities which vary on the relative size of the historically vulnerable population, $Q$:

\begin{table}[h]   \label{Basics of four real cities}
\begin{center}
\begin{minipage}{250pt}
\caption{Demographics of Cities to Generate Synthetic Modeled Cities}\label{tab1}%
\begin{tabular}{@{}lllll@{}}
\toprule
 City &  Population $P$  & Subgroup Q & $Q/P$ &  Number of  \\ [0.5ex] 
 & in 2010 & Population & & Districts\\
\midrule
 Albuquerque & 545,711 & 254,834 & 0.467 & 9\\ 
  &    & Hispanic/Latino& & \\
 Charlotte & 735,847 & 268,404 & 0.365 & 7\\ 

  &    & African-American &  &\\
Pittsburgh &  305,704 & 84,819 & 0.277 & 9\\
 
  &    & African-American & & \\
  Minneapolis & 382,578 & 79,967 & 0.209 & 13\\
 
  &    & African-American & & wards\\
\end{tabular}
\end{minipage}
\end{center}
\end{table}

These populations were calculated by including all people living in all census blocks that intersect any city council district or ward. \cite{USCBP4, USCBP8, USCBSF, ABQS, CHARS, MINNS, PITTS} Note that these are not the populations for Metropolitan Statistical Areas around these cities, because city council or city commission regions generally align with the city boundary and not the larger region. Furthermore, these numbers differ from Census Bureau totals for city populations, because census tract boundaries do not generally align with voting district boundaries.

In selecting cities, several factors were taken into account. We eliminated cities with three or more major populations, in order to simplify the complexity of the task, and to allow the use of the dissimilarity index, $D$. In all four cities, 2010 census data identifies ``Asian alone" as the third largest population after $Q$, and only in the case of Minneapolis is this group large enough (5.63\%) to constitute a voting majority of a single district, and in all cases this third largest group is significantly smaller than the group designated $Q$. This restriction eliminated many cities in California, for example, as well as most cities with total population $P$ above a million people. After that, we sought cities representing a range of ratios between $Q$ and the total population $P$ - roughly speaking, we wished to represent $Q:P$ ratios of 1:2, 1:3, 1:4, and 1:5. We aimed for cities with populations larger than 250K. Finally, where available, we chose cities representing different regions of the United States. (While 2010 data is 12 years out of date, it is no less ``realistic" for our purposes.) 

We let these four cities govern nearly all of the parameters of our synthetic model cities. In other words, a synthetic version of Albuquerque will have the same boundary, same population, same dual graph representation, same regions with dense or sparse housing, and same census block populations as Albuquerque, but the racial compositions of the census blocks will vary to yield an alternate residential segregation pattern. It is as if all residences are left intact, but the people have been redistributed. For each city, 300 synthetic versions are created using a redistribution algorithm to place people into census blocks around the city.   [See Appendix \ref{creating modeled cities} for full details.]

{\bf Note:} The authors wish to acknowledge that this method erases a tremendous amount of history which led to the current racial residence patterns in these actual cities, and we do not trivialize that erasure. In addition, were historians or sociologists to design counterfactual versions of these cities, they would take into account pressures and events that shaped how residential segregation patterns emerge, and many of our synthetic cities would surely seem nonsensical. We hope that a synthetic modeled city merely provides a mathematical comparison for the availability of fair district plans, and does not diminish the lives and history of the actual residents of these cities.

In total, there are 1200 of these synthetic cities, 400 for each of Albuquerque, Charlotte, Pittsburgh, and Minneapolis. We call these synthetic cities our {\em Modeled Cities}, to distinguish them from the rectangular grid cities, in the next section.

\subsubsection{Grid Cities} \label{Grid Cities}

For the remaining 3000 synthetic cities, we build a 30 $\times$30 grid city, with $1000$ individuals living in each census block, for a rectangular city with 900 vertices and 900,000 people. Grid cities are divided into 10 single-member districts.

\begin{figure}[h]
\centering
\includegraphics[scale=.8]{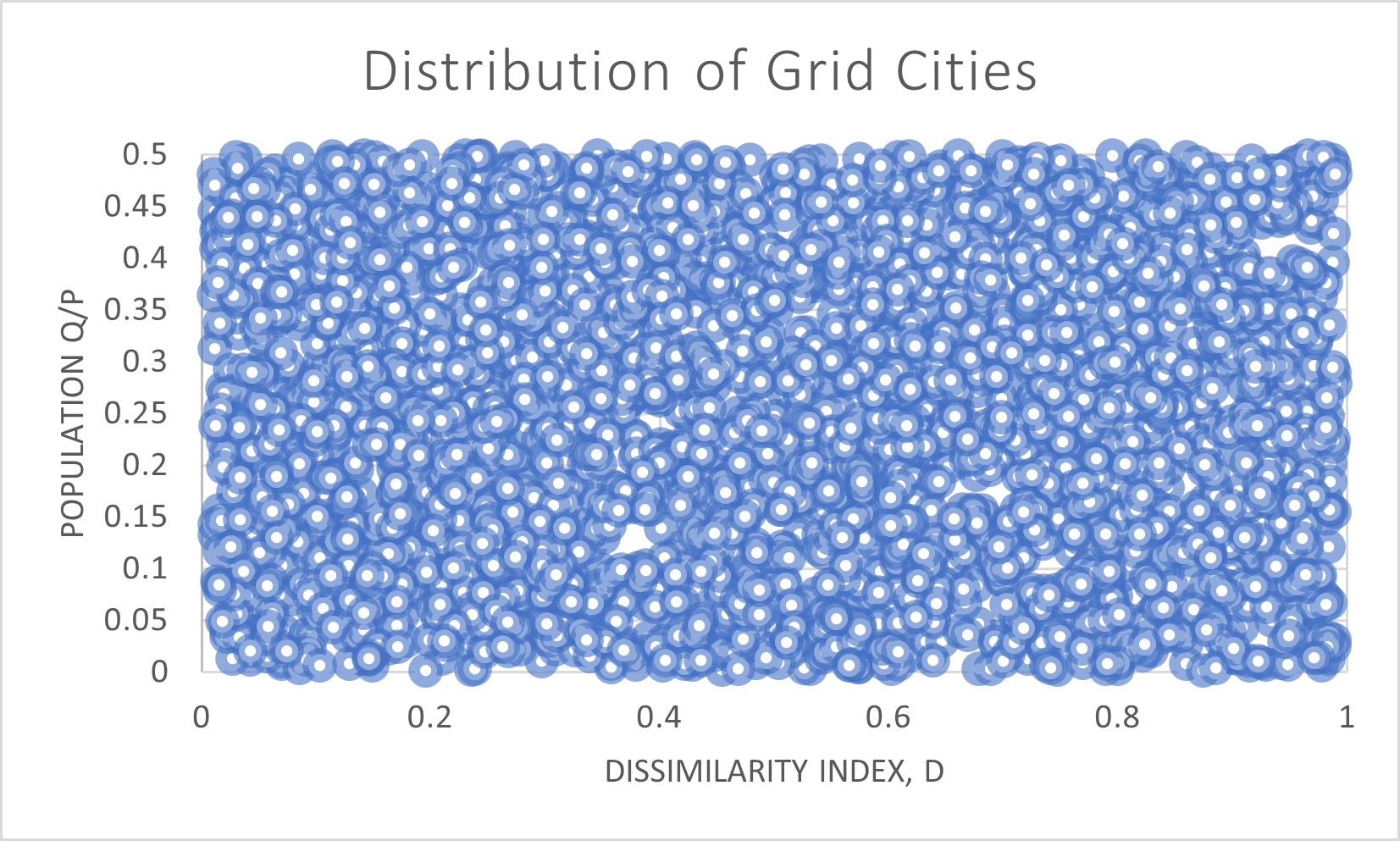}

\caption{Scatter-plot of Grid Cities by population $Q/P$ and dissimilarity index $D$, showing that the sample of cities is uniformly distributed over $0 \leq D \leq 1$ and $0 \leq Q/P \leq 1$}
\label{Distribution of Grid Cities}
\end{figure}

While the modeled cities only differed on their racial residential pattern, the grid cities have two variables of freedom: the size of the minority population, $Q$, and the racial residential pattern.  All possible segregation levels, as measured by the dissimilarity index $D$, and all possible percents $Q$ are represented fairly equally, as shown in Figure \ref{Distribution of Grid Cities}.

The algorithm for placing individuals of $Q$ and $P-Q$ thoughout the city   has asymmetric rules for placing the vulnerable population and the dominant population into vertices. (See Appendix \ref{creating modeled cities}.) When modeling cities based on Albuquerque, Charlotte, Pittsburgh, or Minneapolis, the historically vulnerable population percentage $Q/P$ is always less than $0.5$ of the total population. In creating the grid cities, $Q/P$ is allowed to range from 0 to 1, thus creating cities where the larger population was placed according to the algorithm's rules for the vulnerable population. In this case, we maintain the convention that $Q$ refers to the population with size $< 0.5 P$, regardless of whether it was placed according to the algorithm's rules for vulnerable or dominant populations.

\subsection{Generating the set of district plans and validating the runtime for GerryChain}

We begin with five graphs: the dual graph of Albuquerque, Charlotte, Pittsburgh, Minneapolis, and the 30$\times$30 square grid. For each graph, we began with four seed district plans. For modeled cities, the seeds were the original voting districts, and three others which were generated using the algorithm in Appendix \ref{district plans from scratch}. The grid graph used four generated seeds. For each seed district plan, we then run the software package GerryChain to generate 5000 district plans using GerryChain's recombination proposal. (See \cite{GerryChain} for details on GerryChain, and Appendix \ref{gerrychain settings} for specific settings used here.) This yields 20,000 district plans for each dual graph. 

\subsection{}
{\bf Technique to validate the runtime for GerryChain} \label{validate}

If the sample is only taken from small portion of the statespace, then the statistics of the sample will not be useful evidence for drawing conclusions about the statespace.  Experimental data from 2019 suggests that tens of thousands of ReCom steps in GerryChain will be sufficient to achieve stable results under many real-world scenarios. \cite{DDS19} We seek an additional way to confirm that the algorithm is visiting distant parts of the statespace.
 
 Let $X$ be a city with $n$ districts. Assume that each census block has a constant population and that each district is the union of a constant number of census blocks. Let $\{S_i\}$ be the set of all valid district plans on $X$. Given a random walk through $\{S_i\}$, we would like to assess the spread of the sample throughout the statespace.
 
 Designate some plan as $S_0$, which will function as a quasi-origin in this space. We will describe pairs of district plans as being ``distant" or not, with respect to this reference plan $S_0$.
 
 Each plan $S_i$ is assigned an $n$-tuple ${\bf v_i}$ with values in the set of subsets of census blocks as follows: Let $m_k$ be a district of $S_0$, and consider the intersection of $m_k$ with each of the $n$ districts of $S_i$. Designate the largest intersection as $M_{i,k}$.  (Intuitively, suppose a collection of people live in the district $m_k$, and the city districts are redrawn from $S_0$ to $S_i$. Then $M_{i,k}$ represents the largest group of people from $m_k$ who will continue to vote together in a single district, under the redistricting.)  The largest intersection of $m_k$ need not be unique, in which case any largest will do.
 
 Let ${\bf v_i} = \left< M_{i,1}, M_{i,2}, \dots, M_{i,n} \right>$, the $n$-tuple of sets of census blocks. These entries represent the maximal groups which are preserved if the city $X$ redistricts from $S_0$ to $S_i$. Note that these are not unique - two district plans $S_i$ and $S_j$ could have the same associated $n$-tuple. However, $S_0$ is the only plan which is associated with the $n$-tuple whose entries compose the entire city.

\begin{figure}[h]
    \centering
     \includegraphics[scale=.3]{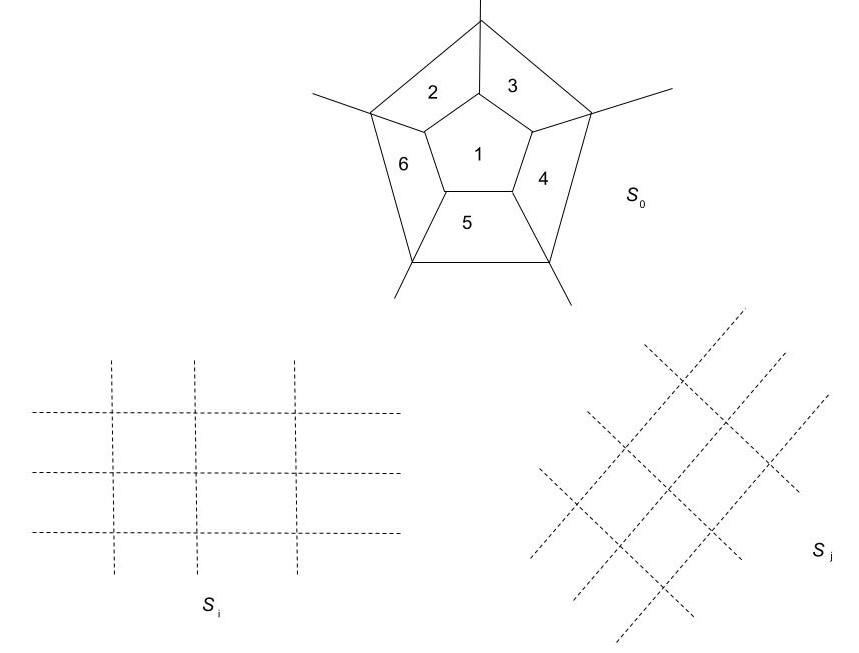}
   \caption{Two district plans, $S_i$ and $S_j$, with a reference plan $S_0$.}
    \label{fig:validation 1}
    \includegraphics[scale=.30]{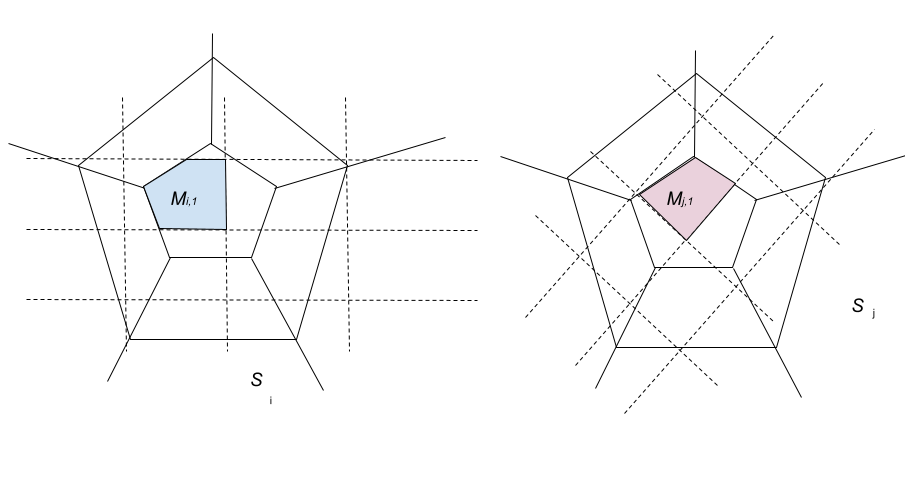}
   \caption{In this figure, $M_{i,1}\cap M_{j,1} \neq \emptyset$, and so $S_i$ and $S_j$ are not distant from each other.}
    \label{fig:validation 2}

    \includegraphics[scale=.35]{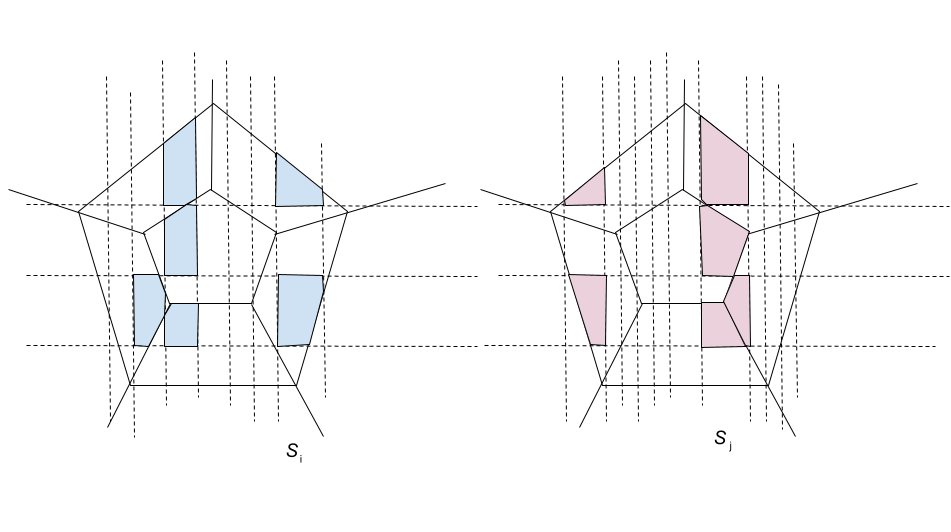}
    \caption{Example of two district plans which are distant with respect to $S_0$.}
    \label{fig:validation 3}
\end{figure}

Given two district plans, $S_i$ and $S_j$, we will say that $S_j$ is {\em distant from} $S_i$ with respect to $S_0$ if $M_{i,k}\cap M_{j,k} = \emptyset$ for all districts $1 \leq i \leq n$ of $S_0$. (This depends so heavily on choice of $S_0$ that it no longer aligns very well with a lay intuition of what similar district plans might mean. At best, we are comparing the disruption caused by redistricting from $S_0$ to $S_i$ to that caused by redistricting from $S_0$ to $S_j$.  If $S_i$ and $S_j$ are distant from each other, then no one is in a maximally preserved group under both redistricting plans.) See Figures \ref{fig:validation 1}, \ref{fig:validation 2}, and \ref{fig:validation 3} for illustration.

The set $M_{i,k}$ was the largest component of $m_k$, after it was fragmented into $\leq n$ components, and so $M_{i,k}$ must contain at least $(1/n)^{th}$ of the census blocks in $m_k$ (by the pigeonhole principle). Therefore, any collection of $\{M_{i,k}\}$ contains at most $n$ mutually disjoint elements.

In addition,  $\cup_k M_{i,k}$ occupies at least $1/n$ of all census blocks in the city. If $S_i$ and $S_j$ are distant from each other with respect to $S_0$, then $M_{i,k}\cap M_{j,k} = \emptyset$ for every district $k$, $1 \leq k \leq n$. Therefore $\cup_k (M_{i,k} \cup M_{j,k})$ occupy at least $2/n$ of all census blocks, and if $S_i, S_j, S_l$ are pairwise all distant from each other, $\cup_k (M_{i,k} \cup M_{j,k} \cup M_{l,k})$ will occupy at least $3/n$ of all census blocks, and so on. Therefore $n$ is an upper bound on the number of district plans which are all pairwise distant from each other. 

We then use this to determine if the sample set $\{S_{i_j}\}$ has explored the statespace $\{S_i\}$. Given the sample $\{S_{i_j}\}$,  we retain a maximal set of mutually distant district plans.  If the size of the set is close to $n$, there is evidence that GerryChain has explored the statespace.  If the size of the set is closer to $1$, it is hard to draw any conclusion about the set $\{S_{i_j}\}$.

\subsubsection{Notes on validation method}

\begin{enumerate}
    \item This notion of ``distant" district plans depends heavily on choice of $S_0$. If $S_0$ contains a long skinny district $m_k$, then intuitively, we may expect that the longer perimeter would increase the number of ways that that $M_{i,k}\cap M_{j,k} = \emptyset$ for two district plans $S_i$ and $S_j$. If $S_0$ contains a district which is roughly square or circular, then the relatively smaller perimeter may yield fewer combinations where $M_{i,k}\cap M_{j,k} = \emptyset$. (In fact, this is borne out by the examples computed, below.) Therefore the same sample set of district plans $\{S_{i_j}\}$ can achieve differing numbers of mutually exclusive distant plans, according to choice of $S_0$.

\item The number of district plans in the statespace is unwieldingly large \cite{MGGG2}, and therefore the number of district plans which are {\em not} distant from $S_i$ with respect to $S_0$ is extremely large. Thus this is a very coarse division of our statespace. We can roughly say how many large, rambling territories of the statespace we have visited, but this is not a very detailed claim. Still, it does provide evidence that we are not trapped in a small region of the statespace.

\item We have assumed that each district contained the same number of census blocks, and that each census block contained the same number of people. This facilitated an intuitive notion of $M_{i,k}$: the largest block of people in district $k$ who would still vote together under redistricting from $S_0$ to $S_i$.  

These assumptions hold for our grid cities, but do not hold for the modeled cities; census blocks can range from zero to hundreds of people. Since districts must be kept roughly similar in population, the number of census blocks in a district is therefore highly variable.  Each $M_{i,k}$ still occupies at least $1/n$ of the census blocks in district $k$, although the number of census blocks in each district is no longer constant. However, it is still true that $n$ is an upper bound on the number of distant district plans.
\end{enumerate}
On the plus side, we note that the this computation is relatively quick to implement. 

\subsection{Implementation of GerryChain Validation}

\subsubsection{Grid Cities}

Grid cities have 10 districts, and so the maximum number of mutually distant district plans is 10.  In this case, we chose three different seeds $S_0$, and found a maximally large set of mutually distant plans within our 20,000 district plans with respect to each seed.

The first seed was a randomly generated choice of $S_0$. (See Appendix \ref{district plans from scratch}.) Under this choice, the 20,000 district plans yielded a set of 7 mutually distant district plans. The second choice of seed was a rectangular grid pattern of 10 districts. While $n=10$ districts does not lend itself to a checkerboard pattern, the attempt was to get reasonably close to evenly square districts. Out of a possible 10, there were again 7 mutually distant district plans with respect to this seed. The final choice of seed was a pattern of 10 horizontal  rectangles, each running the full width of the city. Given two district plans $S_i$ and $S_j$, we expect that it is much easier for $M_{i,k} \cap M_{j,k} = \emptyset$ with respect to this $S_0$. And in fact, when the 20,000 district plans are checked with respect to this seed, there are 9 mutually distant district plans.

Given this information, we conclude that we have reasonable evidence that the random walk is in fact moving throughout the statespace of district plans, and not restricted to a small portion.

\subsubsection{Modeled Cities}

Albuquerque, Charlotte, Pittsburgh, and Minneapolis have 9 districts, 7 districts, 9 districts, and 13 wards respectively. Therefore the  maximum number of mutually distant district plans for each city is 9, 7, 9, and 13, in kind.

For each city, $S_0$ was created randomly. (See Appendix \ref{district plans from scratch}). The results are summarized in Table \ref{Validation of Modeled Cities}.

\begin{table}[h]  
\begin{center}
\begin{minipage}{300pt}
\caption{Demographics of Cities to Generate Synthetic Modeled Cities}
\label{Validation of Modeled Cities}
\begin{tabular}{@{}llll@{}}  
\toprule
  City &  Sample size  & Number of  mutually&  Upper bound on   \\ 
   &   & distant plans &   mutually distant plans  \\
\midrule
  Albuquerque & 20,000 & 7 & 9 \\ 
  Charlotte & 20,000  & 6 &  7 \\ 

 Pittsburgh &  20,000  & 7 & 9\\

  Minneapolis & 20,000  & 7 & 13\\
\end{tabular}
\end{minipage}
\end{center}
\end{table}

Thus we have evidence that the random walk was not constrained to a small portion of the statespace of district plans.  

\section{Results} \label{results}

In this section, we demonstrate the correlation between segregation and the availability of a fair district plan.
  
\subsection{Synthetic Modeled Cities}

Recall that each real city has 300 synthetic cities modeled on it. For each synthetic modeled city, we have computed its measure of segregation, $D$, and measure of Fairness, $\bar F$. The dissimilarity index $D$ indicates how closely the racial proportions of each census block match the city-wide racial proportions. A value of $D=1$ indicates that every census tract is either entirely composed of members of $Q$ or has no one from $Q$, and $D=0$ indicates that the ratio $Q/P$ within every census tract is the same as the citywide ratio $Q/P$.  The fairness index, $\bar F$, is intended to describe availability of a district plan with a proportional number of minority-majority district, within the set of sampled district plans. It is computed by taking the average percent of minority-majority districts across the ensemble of district plans, and then normalizing the average by dividing by $Q/P$.
 A value of $\bar F = 1$ indicates that, across the sample, the average percent of minority-majority districts is equal to $Q/P$. When  $\bar F < 1$, the vulnerable population on average will have fewer minority-majority districts, proportionally, than their share of the population, and if $\bar F > 1$, they will have more minority-majority districts, proportionately.

\subsubsection{Synthetic Albuquerque cities}

In 2010, Albuquerque had a population of 545,711 people with 254,834 identified as Hispanic-Latino by the census, and 9 city council districts. Therefore $Q/P = 0.467$ in our model Albuquerque cities, and it takes 30,317 people to constitute a majority of a district.   There are 9,149 census blocks, and thus vertices in the dual graph. A fair district plan would then yield between 4-5 minority-majority district plans, and $\bar F=1$ would indicate that on average, a city has 4.20 minority-majority districts across all 20,000 district plans.

\begin{figure}[h]
    \centering
    \includegraphics[scale=.75]{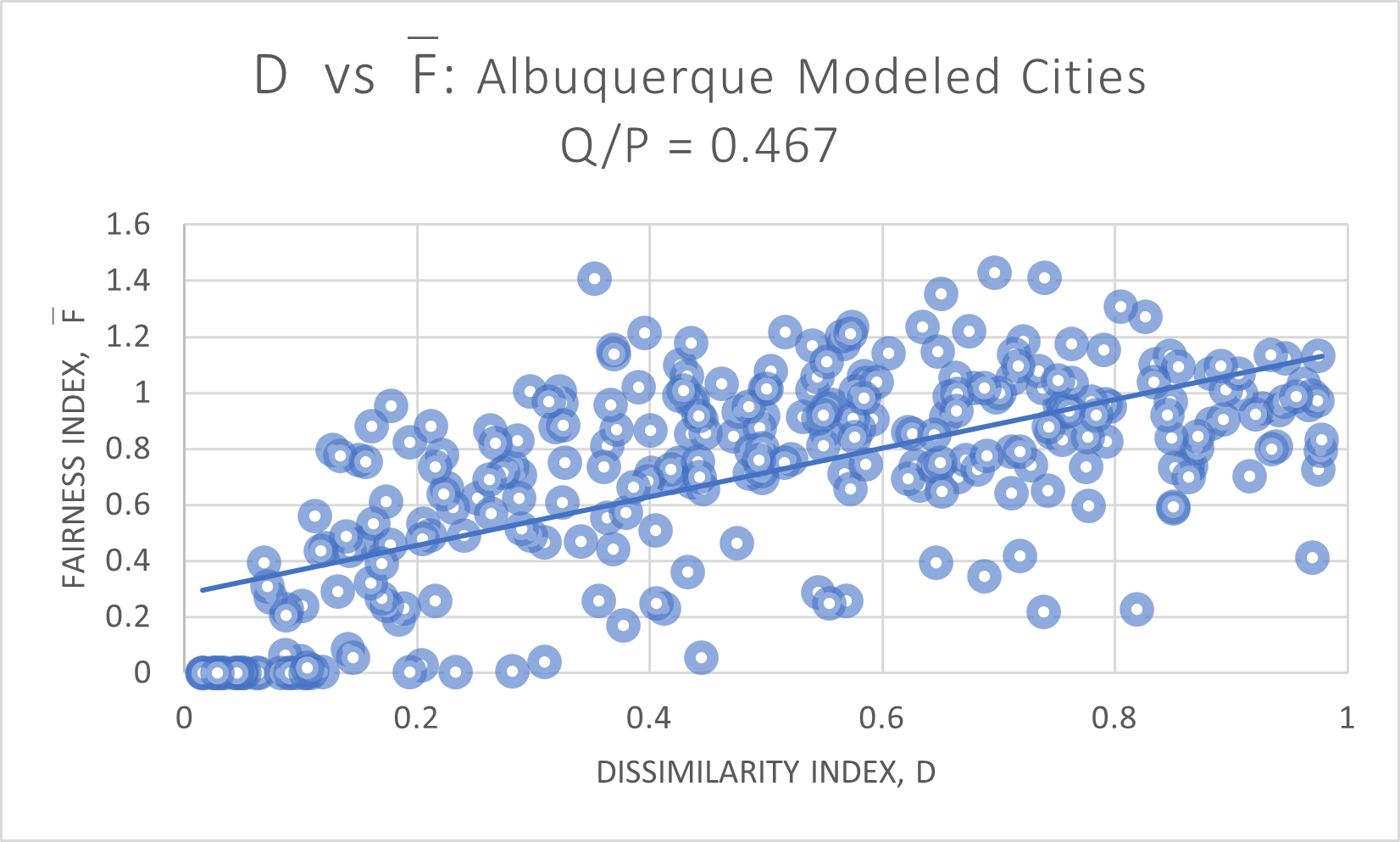}
   \caption{Scatter-plot of Albuquerque Modeled Cities by fairness index, $\bar F$ and dissimilarity index $D$, with trendline}
    \label{D vs F: Alb Modeled Cities}
\end{figure}

The 300 synthetic Albuquerque cities exihibit a modest degree of correlation between residential segregation and availability of a fair district plan, shown in Figure \ref{D vs F: Alb Modeled Cities}.

The slope of the regression line is 0.8664, and $R^2 = 0.427$. No matter how the residents are distributed about the city, there is often a fair district map available, although we note that as $D$ approaches 0.1, it becomes increasingly hard to find proportional district plans.

(We do wish to note once more that in the real world, districts often have a lower threshold for qualifying as opportunity districts than being strictly  minority-majority, as we have defined here. Political scientists determine the threshold for opportunity districts by measuring voter turnout, degree of racially polarized voting, community activism, and other nuanced characteristics of a specific group of people.)\\

\subsubsection{Synthetic Charlotte cities}

In 2010, Charlotte had a population of 735,847 people with 268,404 identified as African-American by the census, and 7 city council districts. Therefore $Q = 36.5\%$ in our model Charlotte cities, and it takes 52,561 people to constitute a majority of a district.   There are 8,822 census blocks, and thus vertices in the dual graph. A fair district plan would then yield between 2-3 minority-majority district plans, and $\bar F=1$ would indicate that on average, a city has 2.56 minority-majority districts across all 20,000 district plans.

In Figure \ref{D vs F: Char Modeled Cities} we see a strong correlation between segregation and opportunity districts:
 
 \begin{figure}[h]
    \centering
     \includegraphics[scale=.75]{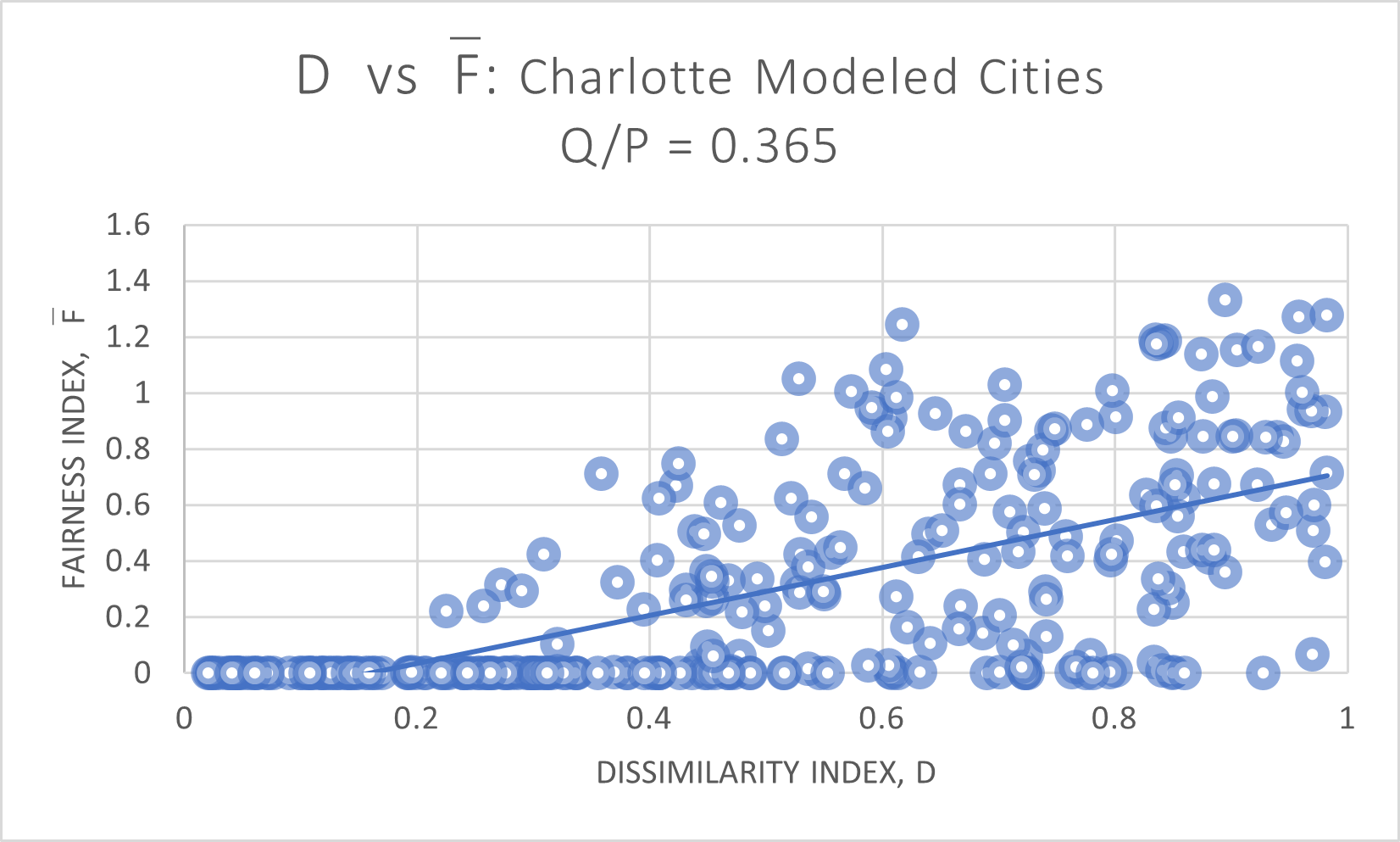}
   \caption{Scatter-plot of Charlotte Modeled Cities by fairness index, $\bar F$ and dissimilarity index $D$, with trendline}
    \label{D vs F: Char Modeled Cities}
\end{figure}
 
The slope of the regression line is 0.8554, and correlation coefficient $R^2 = 0.4245$.  Lower values of $D$ indicate a more diffuse, even distribution of the population $Q$. In many of these cities, it is very hard to achieve even one opportunity district. In other words, while $Q$ is more than five times larger than the threshold to win a single district, these populations are spread so evenly as to be self-cracking - it is impossible to draw a map which gives a 2-3 minority-majority districts. This is a demonstration of the phenomenon described in {\em Locating the Representational Baseline: Republicans in Massachusetts} \cite{DGHKNW19}. 

In contrast, when synthetic cities have high $D$,  far more of the $Q$ population lives in census blocks which are more homogeneous than the citywide average.  In these cities, maps with proportional-opportunities are far more likely to be available. 

\subsubsection{Synthetic Pittsburgh cities}

\begin{figure}[h]
    \centering
    \includegraphics[scale=.75]{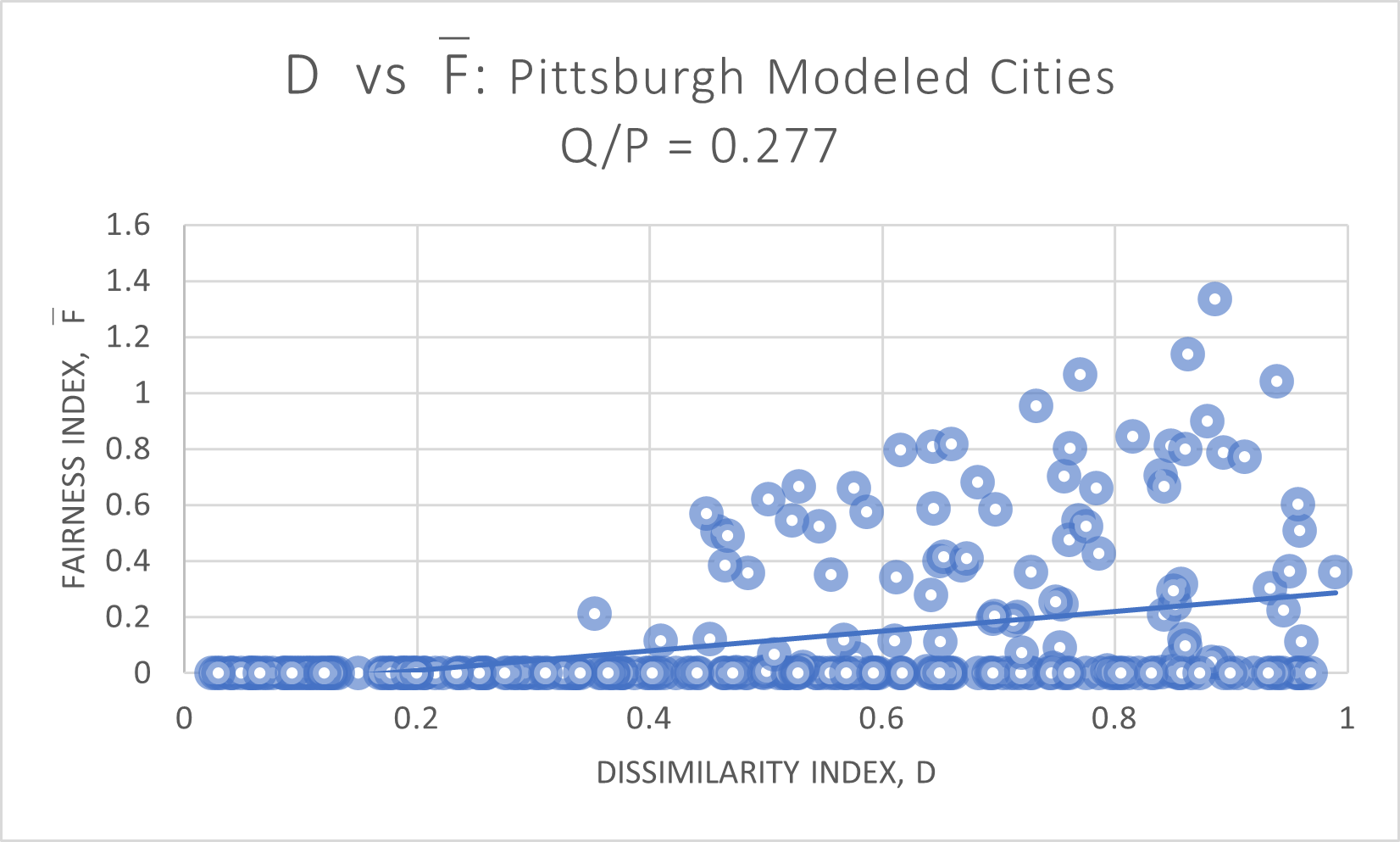}
   \caption{Scatter-plot of Pittsburgh Modeled Cities by fairness index, $\bar F$ and dissimilarity index $D$, with trendline}
    \label{D vs F: Pitt Modeled Cities}
\end{figure}

In 2010, Pittsburgh had a population of 305,704 people with 84,819 identified as African-American by the census, and 9 city council districts. Therefore $Q = 27.7\%$ in our model Pittsburgh cities, and it takes 16,984 people to constitute a majority of a district.   There are 8,822 census blocks, and thus vertices in the dual graph. A fair district plan would then yield between 2-3 minority-majority district plans, and $\bar F=1$ would indicate that on average, a city has 2.49 minority-majority districts across all 20,000 district plans. (See Figure \ref{D vs F: Pitt Modeled Cities}.)

The regression line has slope of 0.3515, and correlation coefficient $R^2$ is  0.1523. For cities with $D < 0.6$, it is rare that a proportional district plan is easily available. Below to $D = 0.4$, there are many cities where it is virtually impossible to find a single minority-majority district.

\subsubsection{Synthetic Minneapolis cities}

\begin{figure}[h]
    \centering
     \includegraphics[scale=.75]{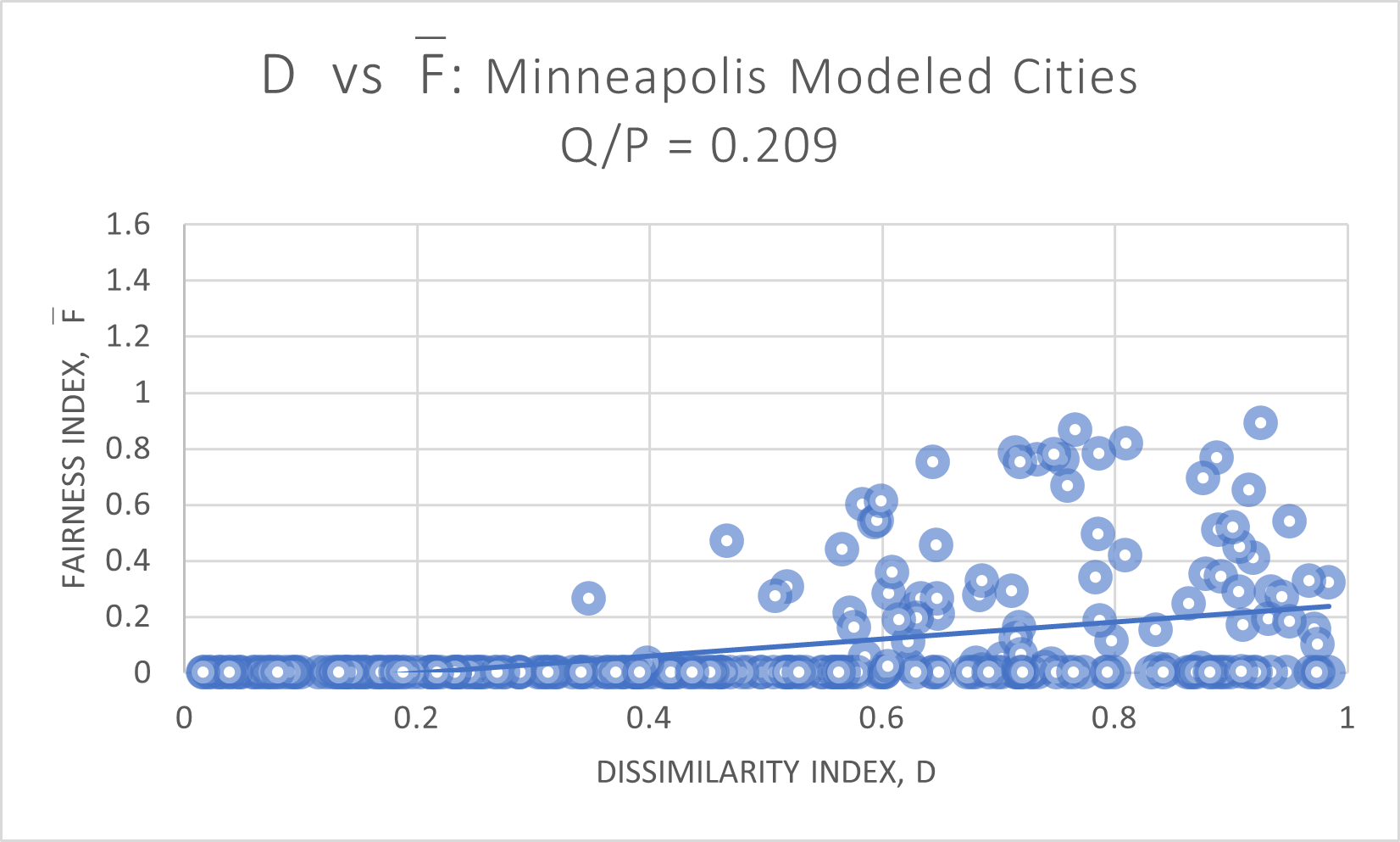}
   \caption{Scatter-plot of Minneapolis Modeled Cities by fairness index, $\bar F$ and dissimilarity index $D$, with trendline}
    \label{D vs F: Minn Modeled Cities}
\end{figure}

Finally, Minneapolis had a population of 382,578 on the 2010 census, with 79,967 people identified as African-American, for $Q=20.9\%$. With 13 wards, a majority in any district is 3.85\% of the population and requires 14,715 people. A proportional district plan yields 2-3 minority-majority wards, and $\bar F=1$ would indicate that an average city has 2.72 minority-majority wards, on average. Results are shown in Figure \ref{D vs F: Minn Modeled Cities}.

Here, the regression line has slope 0.3016 and $R^2 = 0.1798$. As before, as $D$ tends towards 0.6 or lower, it becomes impossible to locate even a single minority-majority ward.

\subsection{Grid Cities}

Our rectangular grid cities vary on three properties: percent minority population $Q$, dissimilarity index $D$, and fairness index $\bar F$.  Populations live along a $30 \times 30$ rectangular grid, with uniform population distribution of 1000 people per vertex. Only the segregation patterns are varied, which amounts to varying the percent $Q$ at each vertex.

\subsubsection{Grid Cities Sometimes Resemble Synthetic Cities}

As noted in Section \ref{Grid Cities}, the full population of all 3000 grid cities are spread out evenly across all possible segregation levels $0 \leq D \leq 1$ and population $0 \leq Q/P \leq 0.5$. (In Figure \ref{Distribution of Grid Cities, 2}, we re-print Figure \ref{Distribution of Grid Cities} for reference.)

\begin{figure}[h]
\centering
\includegraphics[scale=.8]{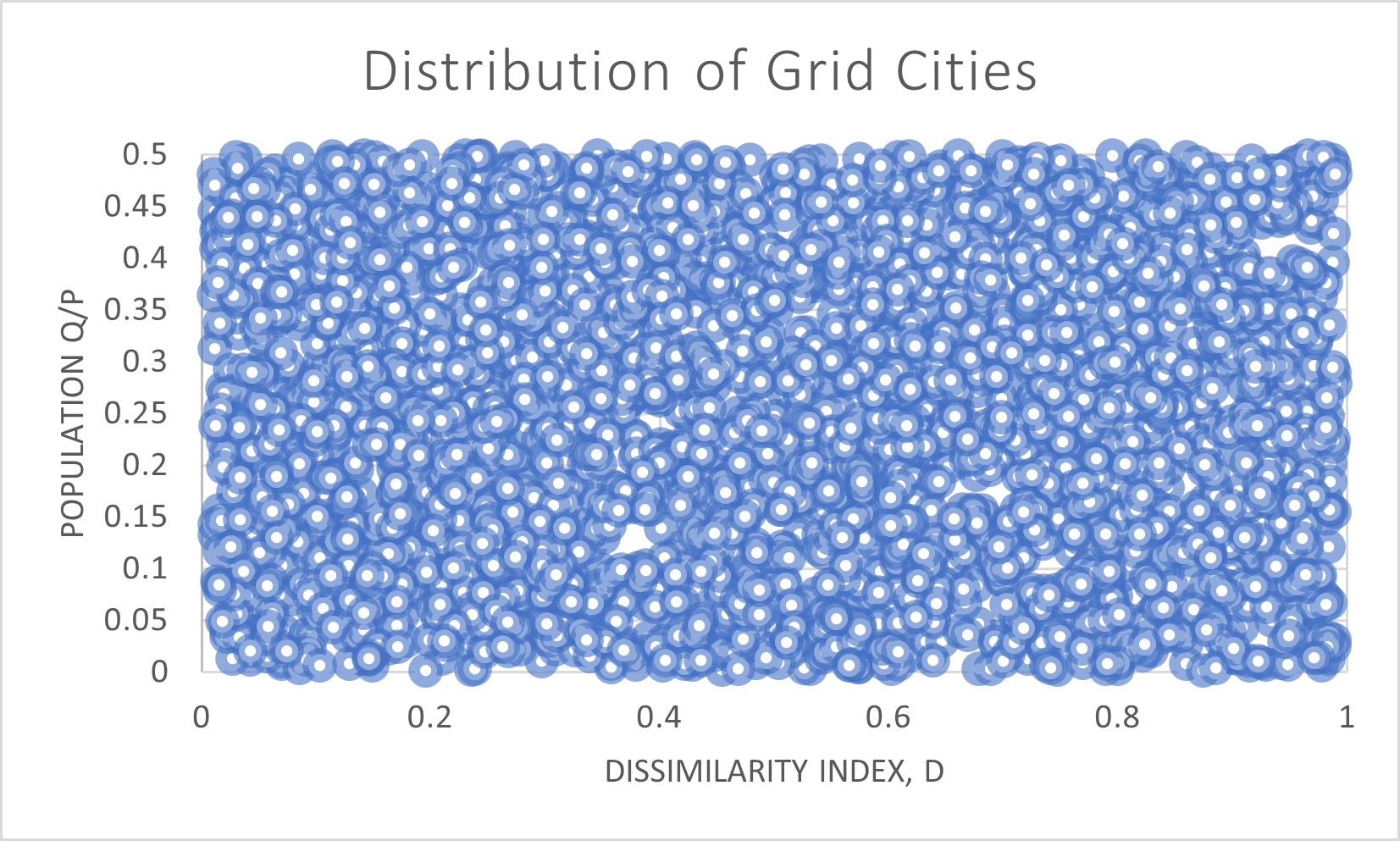}

\caption{Scatter-plot of grid cities by population $Q/P$ and dissimilarity index $D$, showing that the sample of cities is uniformly distributed over $0 \leq D \leq 1$ and $0 \leq Q/P \leq 1$ (shown earlier as Figure \ref{Distribution of Grid Cities}).}
\label{Distribution of Grid Cities, 2}
\end{figure}

We first demonstrate that the extent to which the grid cities behave similarly to the non-grid cities.  In short, the general trendlines are similar, but on the whole, grid cities generally yield higher $\bar F$ than the modeled cities.

\begin{figure}[h]
\centering
\includegraphics[scale=.45]{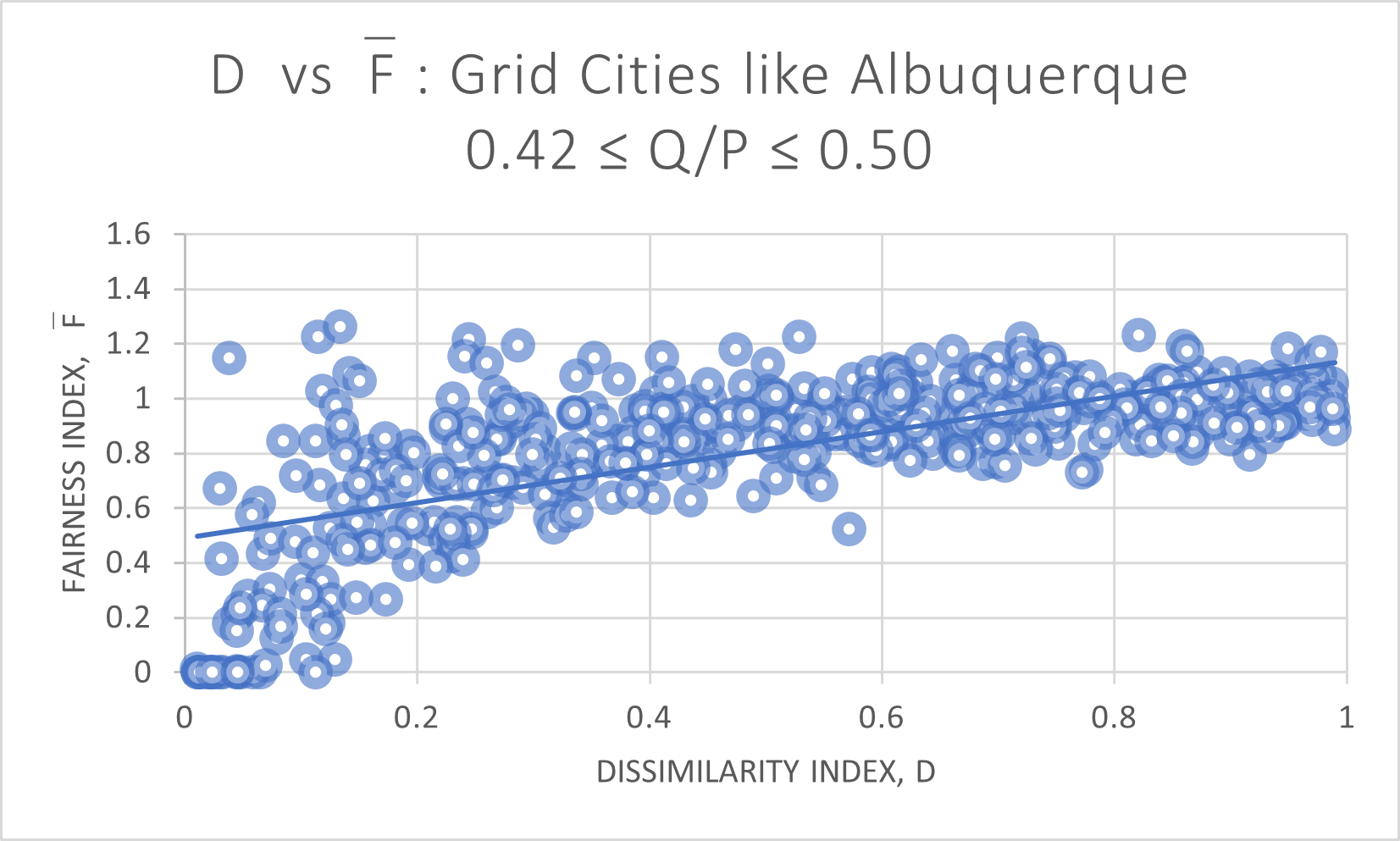}
\includegraphics[scale=.45]{ABQ_D_v_F.png}

\caption{Two scatter-plots of cities by fairness index $\bar F$ and dissimilarity index $D$, for comparison. The first set is of Grid Cities with $0.42 \leq Q/P \leq 0.50$ and the second set is of the Albuquerque modeled cities.}
\label{Grid Cities like ABQ}
\end{figure}

\begin{figure}[h]
\centering
\includegraphics[scale=.45]{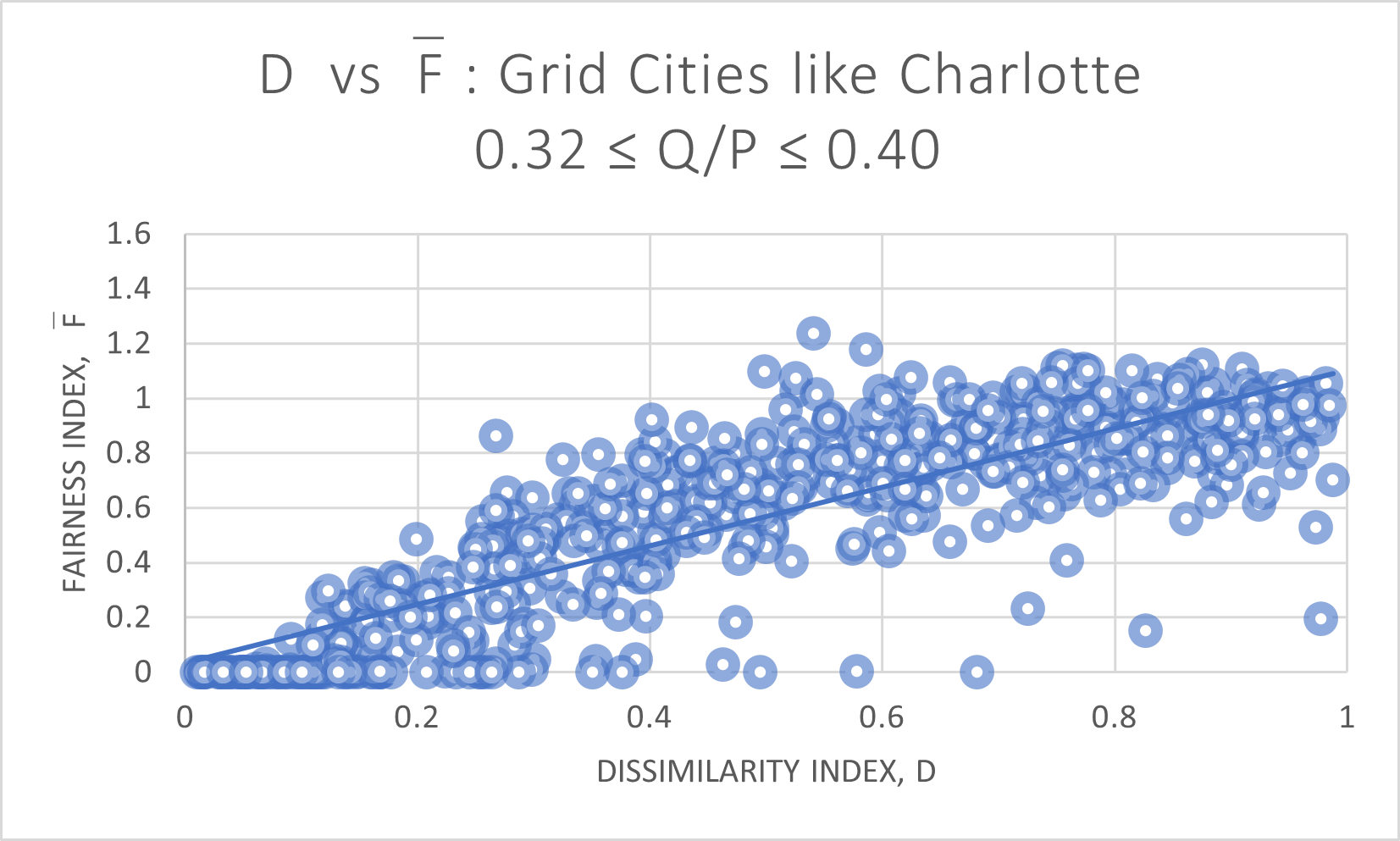}
\includegraphics[scale=.45]{CHAR_D_v_F.png}

\caption{Two scatter-plots of cities by fairness index $\bar F$ and dissimilarity index $D$, for comparison. The first set is of Grid Cities with $0.32 \leq Q/P \leq 0.40$ and the second set is of the Charlotte modeled cities.} 
\label{Grid Cities like Char}
\end{figure}

\begin{figure}[h]
\centering
\includegraphics[scale=.45]{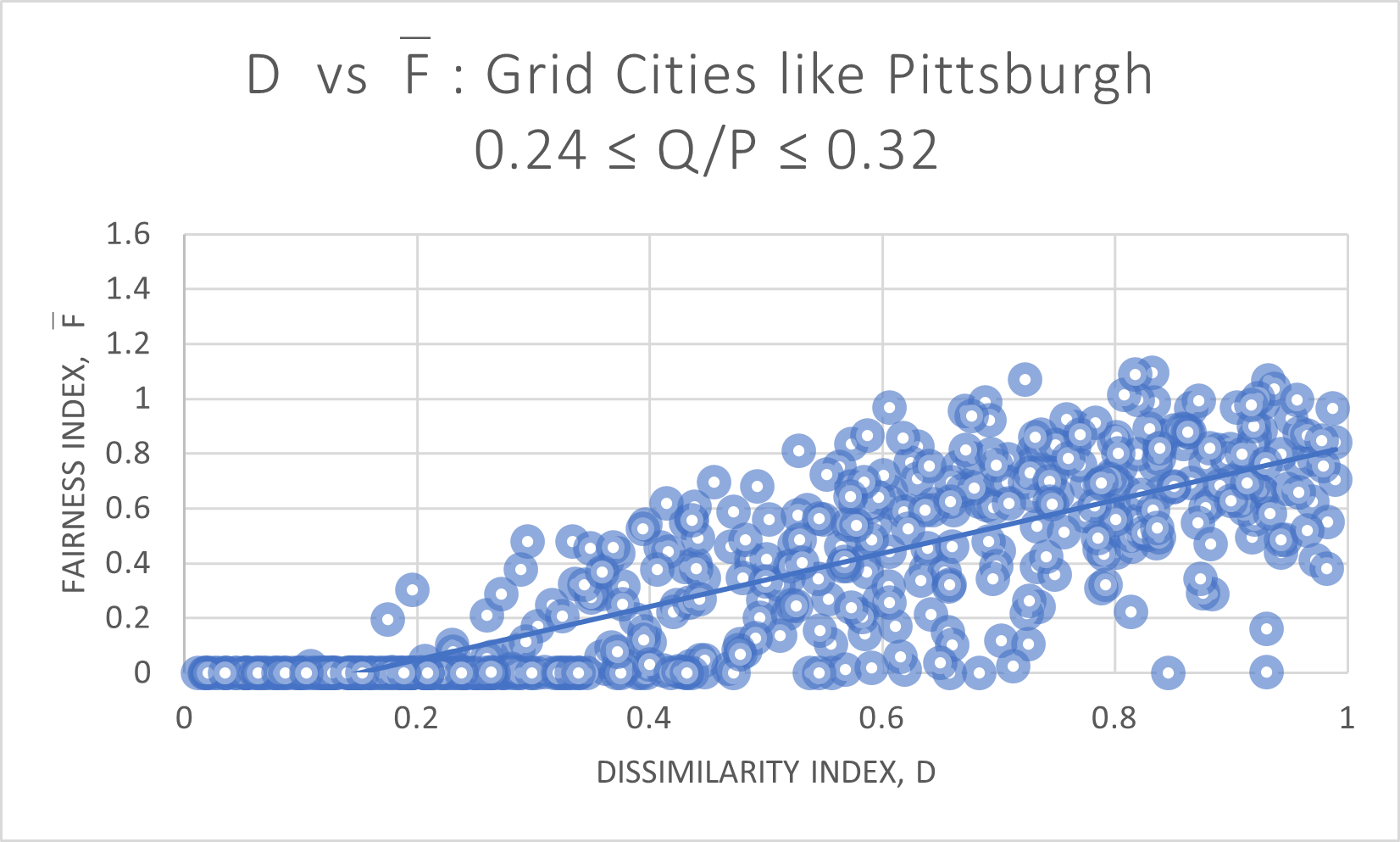}
\includegraphics[scale=.45]{PITT_D_v_F.png}

\caption{Two scatter-plots of cities by fairness index $\bar F$ and dissimilarity index $D$, for comparison. The first set is of Grid Cities with $0.24 \leq Q/P \leq 0.32$ and the second set is of the Pittsburgh modeled cities.} 
\label{Grid Cities like Pitt}
\end{figure}

\begin{figure}[h]
\centering
\includegraphics[scale=.45]{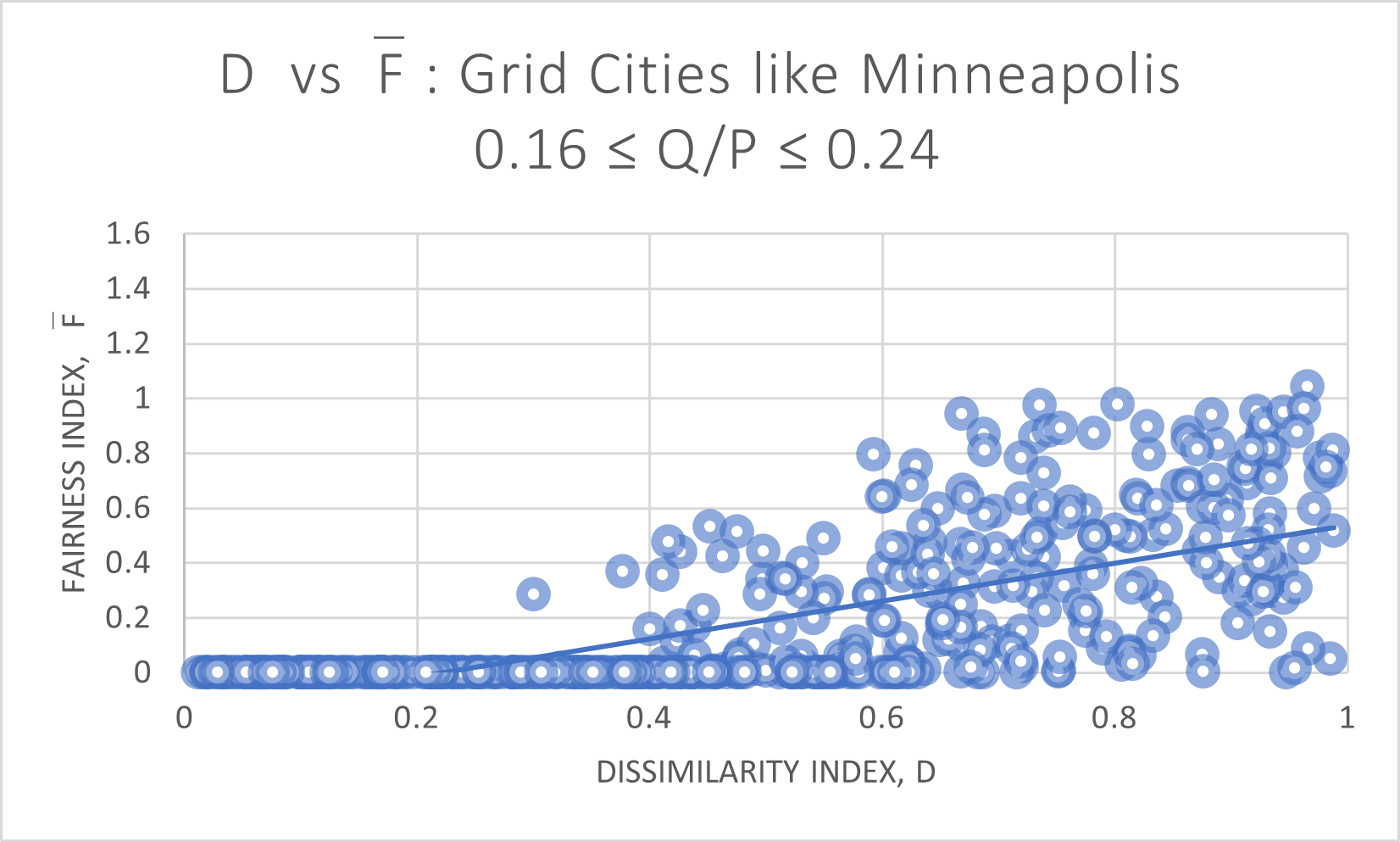}
\includegraphics[scale=.45]{MINN_D_v_F.png}

\caption{Two scatter-plots of cities by fairness index $\bar F$ and dissimilarity index $D$, for comparison. The first set is of Grid Cities with $0.16 \leq Q/P \leq 0.24$ and the second set is of the Minneapolis modeled cities.} 
\label{Grid Cities like Minn}
\end{figure}

Recall that Albuquerque is 0.467 Hispanic/Latino. In Figure \ref{Grid Cities like ABQ}, we restrict our attention to those grid cities with $0.42 \leq Q/P \leq 0.5$. 
 The slope of the trendline on the grid cities in this range is 0.6463, with an average $\bar F = 0.825$. In comparison, the modeled Albuquerque cities have a trendline with slope of 0.8664 and average $\bar F = 0.702$. For Charlotte, we compare grid cities with $0.32 \leq Q/P \leq 0.4$, in Figure \ref{Grid Cities like Char}.
Here, the grid cities trendline has a slope of 1.0681 and average $\bar F = 0.565$, compared to 0.8554 slope and average $\bar F = 0.289$ for the modeled Charlotte cities. For the comparison to Pittsburgh modeled cities, we consider grid cities with $0.24 \leq Q/P \leq 0.32$ in Figure \ref{Grid Cities like Pitt}.
The regression line for the grid cities has slope 0.9733 and average $\bar F = 0.349$, while the modeled Pittsburgh cities have a line with slope 0.3515 and average $\bar F = 0.118$. Finally, with Minneapolis modeled cities, we consider grid cities where $0.16 \leq Q/P \leq 0.24$ in Figure \ref{Grid Cities like Minn}.
The grid cities in this range have a trendline slope of 0.6884 and average $\bar F = 0.187$, while the modeled cities have a slope of 0.3016 and average $\bar F = 0.089$.

Clearly the modeled cities only bear a partial resemblance to the distributions of the modeled cities. We are unable to say whether the grid cities or modelled cities are more meaningful for modeling the possibilities of actual cities. Both are fictitious and have their limitations. 

Our hope is that if we restrict our attention to those circumstances where these different methodologies yield compatible results, then it is more likely that we are capturing something meaningful about actual cities in the real world.

\subsubsection{Dissimilarity Index vs. Percent Minority}

We can then extrapolate along different conditions. For a fixed Fairness, which combinations of $D$ and $P/Q$ achieve $\bar F$ in this range? \\

\begin{figure}[h]
\centering
 \includegraphics[scale=.5]{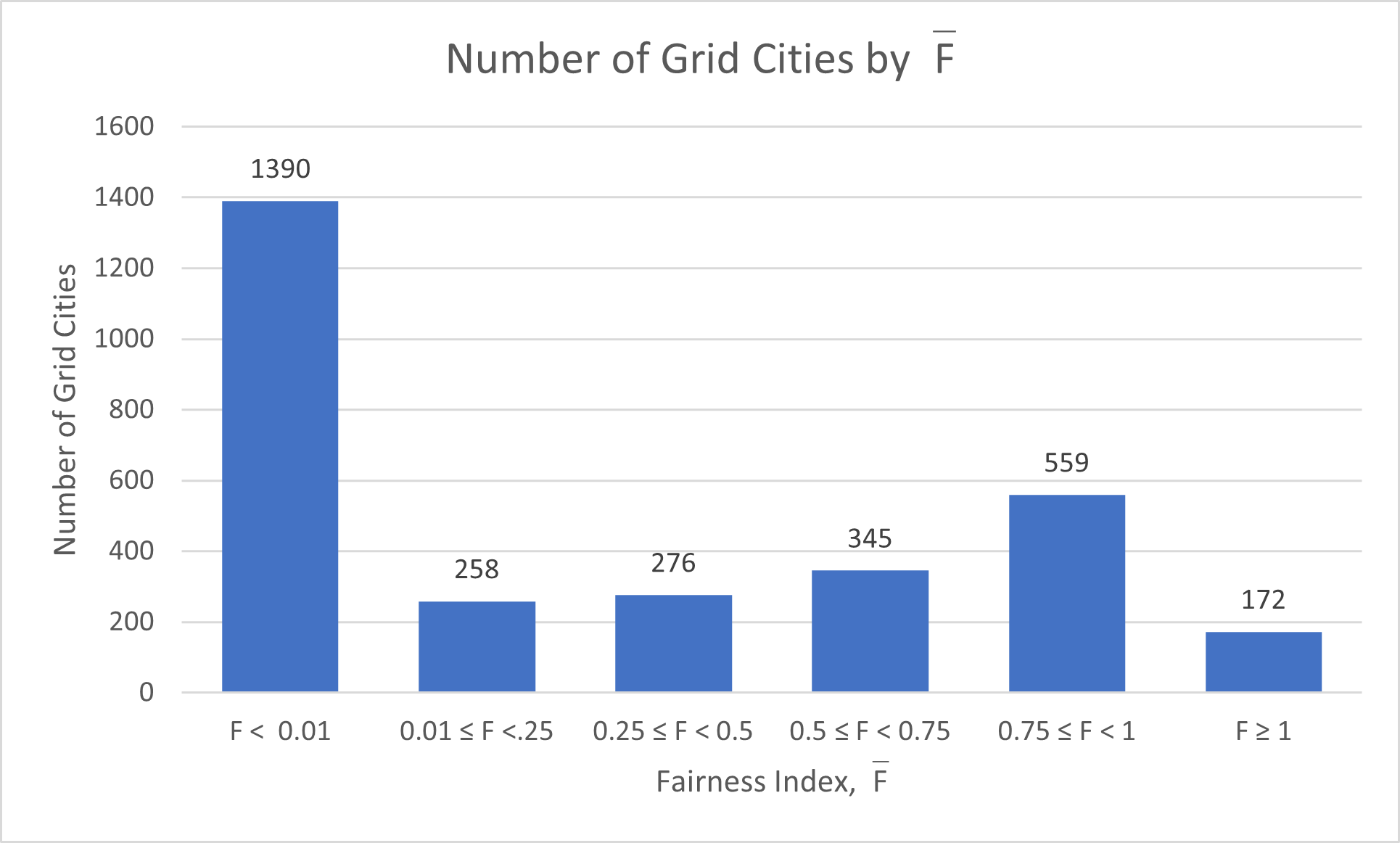}

\caption{A histogram of Grid Cities by fairness index $\bar F$, showing 1390 cities with $\bar F < 0.01$} 
\label{Histogram of Grid Cities by F}
\end{figure}

We first note that the vast majority of grid cities score very low on $\bar F$, as shown in Figure \ref{Histogram of Grid Cities by F}, with a full 1390 of the 3000 cities in the negligibly small $\bar F < 0.01$ category. 
The vast majority of district plans in these cities do not contain a single available minority-majority district in any district plan.\\

\begin{figure}[h]
\centering
 \includegraphics[scale=.6]{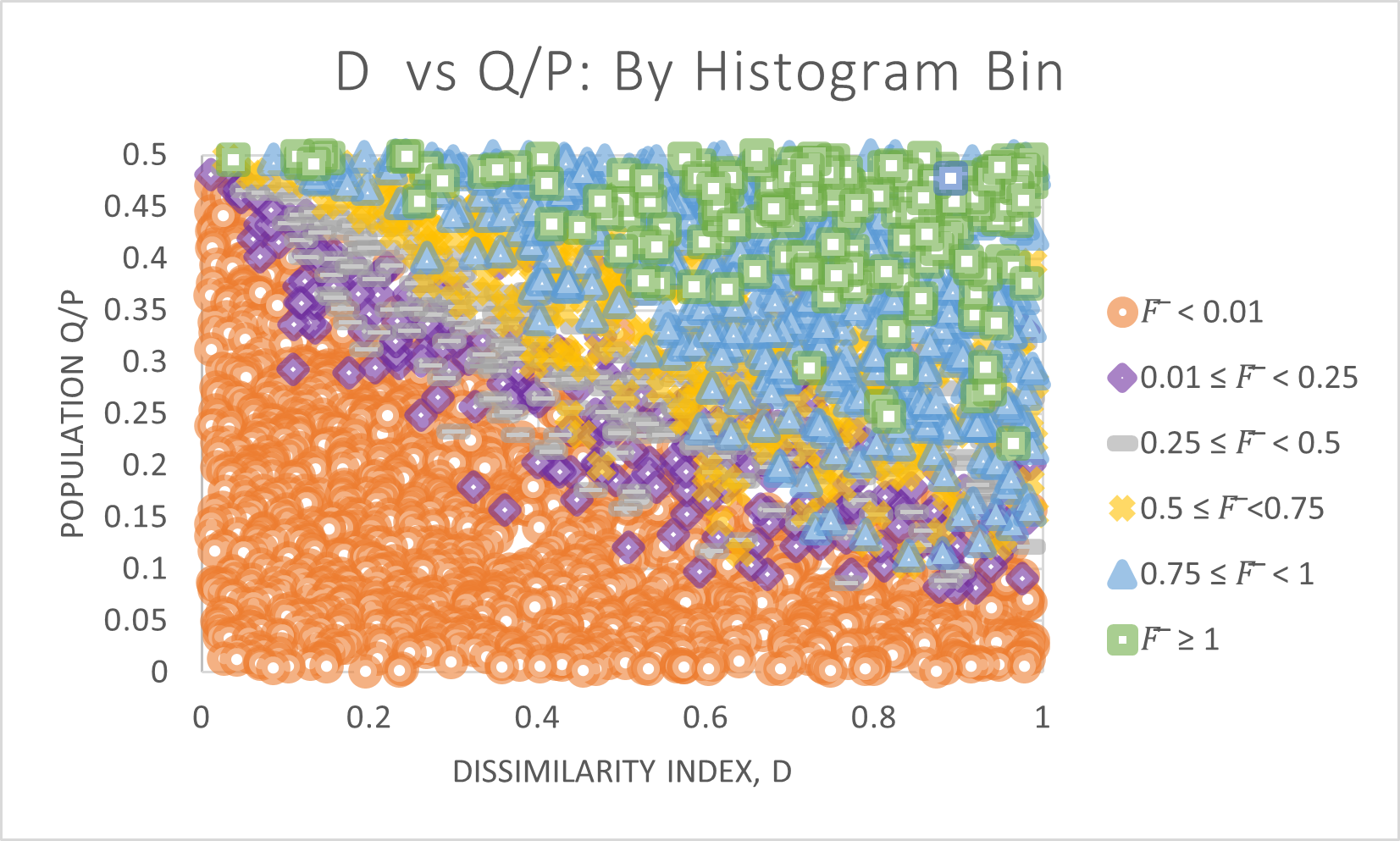}
\caption{Scatter-plot of Grid Cities by population $Q/P$ and dissimilarity index $D$, with colors showing each $\bar F$ histogram bin} 
\label{Scatter-plot Grid Cities by histogram bin}
\end{figure}

The histogram bins can each be individually highlighted, out of the full 3000 grid cities, as shown in Figure \ref{Scatter-plot Grid Cities by histogram bin}.

\begin{figure}[h]
\centering
\includegraphics[scale=.5]{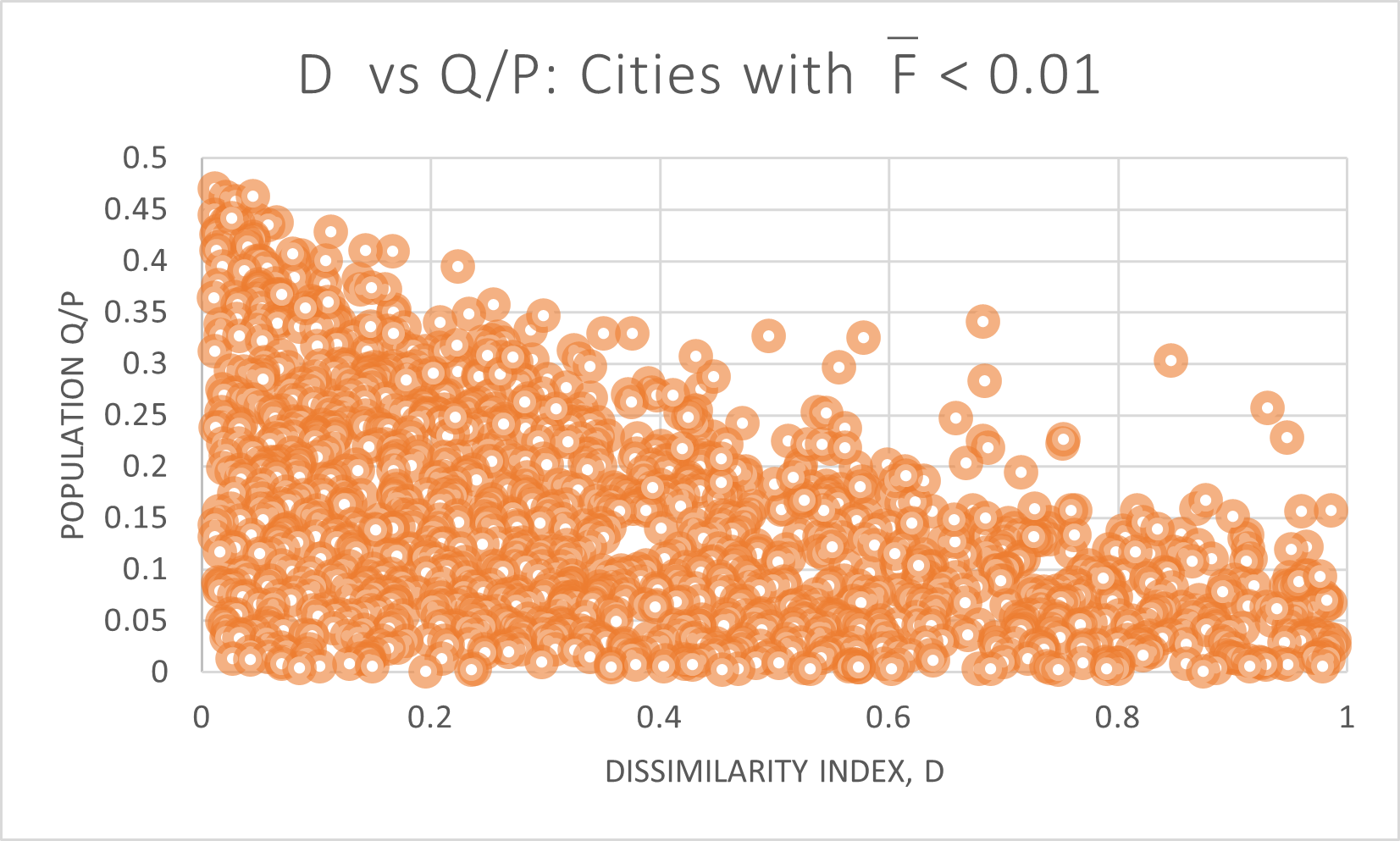}
\includegraphics[scale=.5]{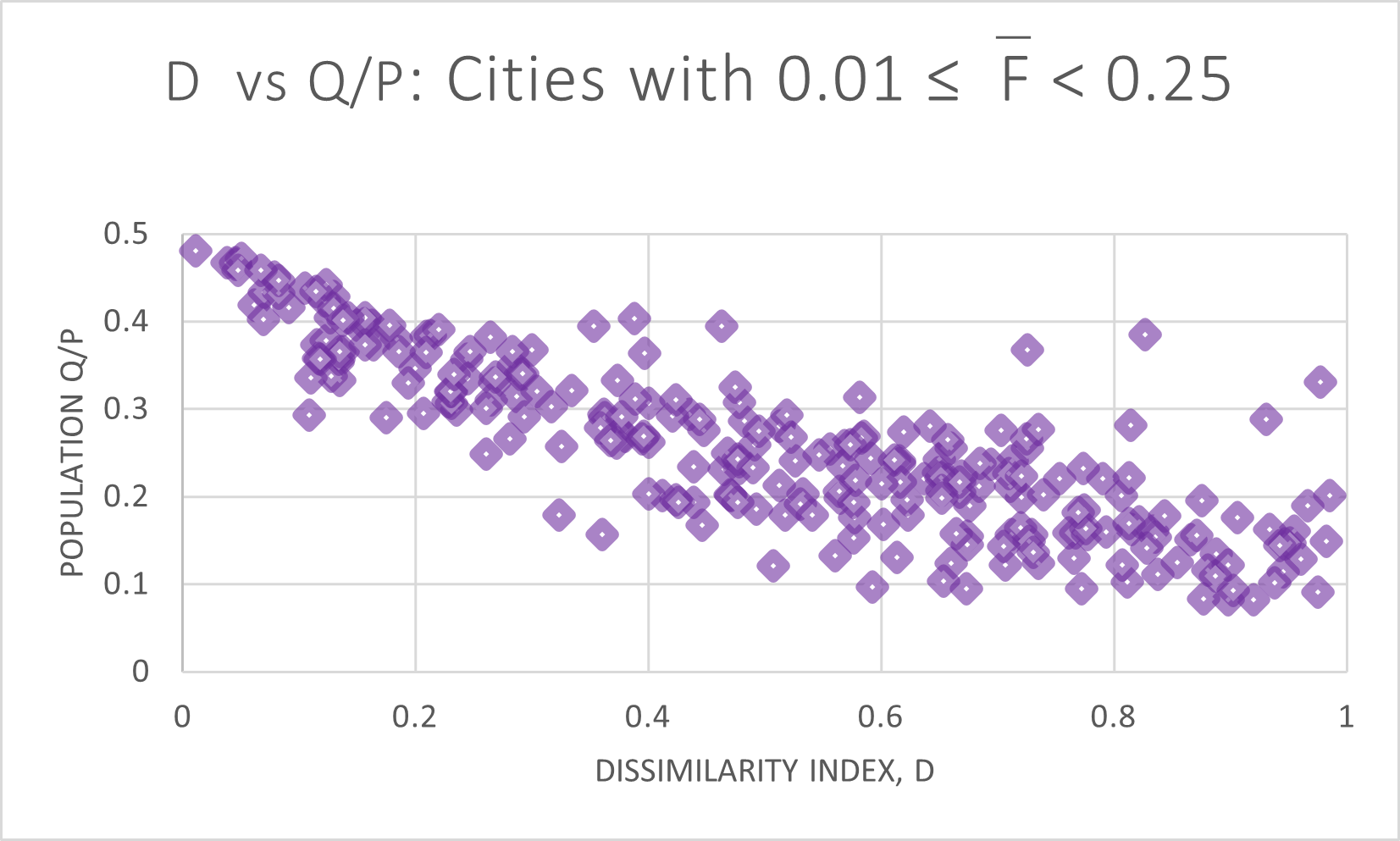}
\includegraphics[scale=.5]{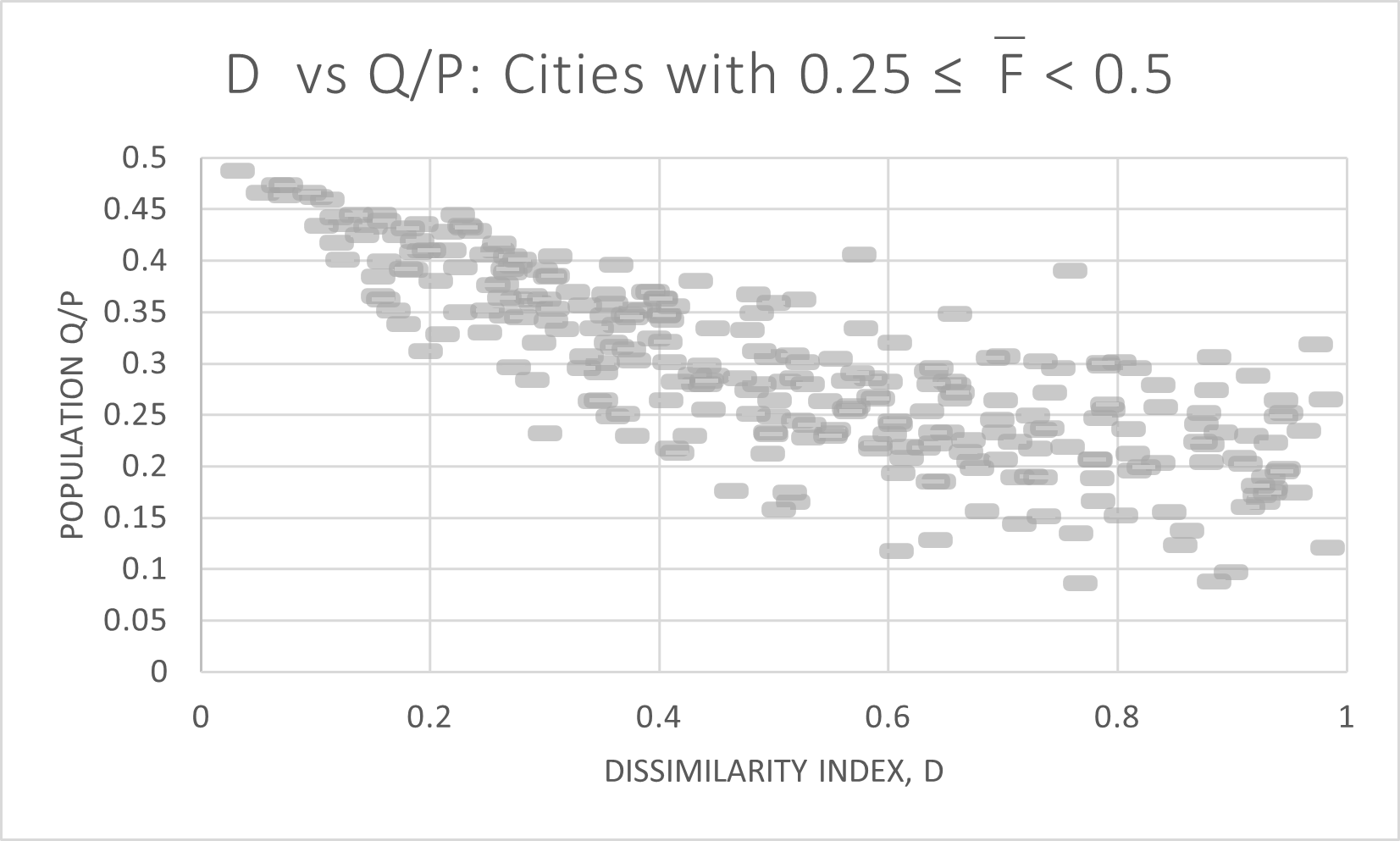}
\includegraphics[scale=.5]{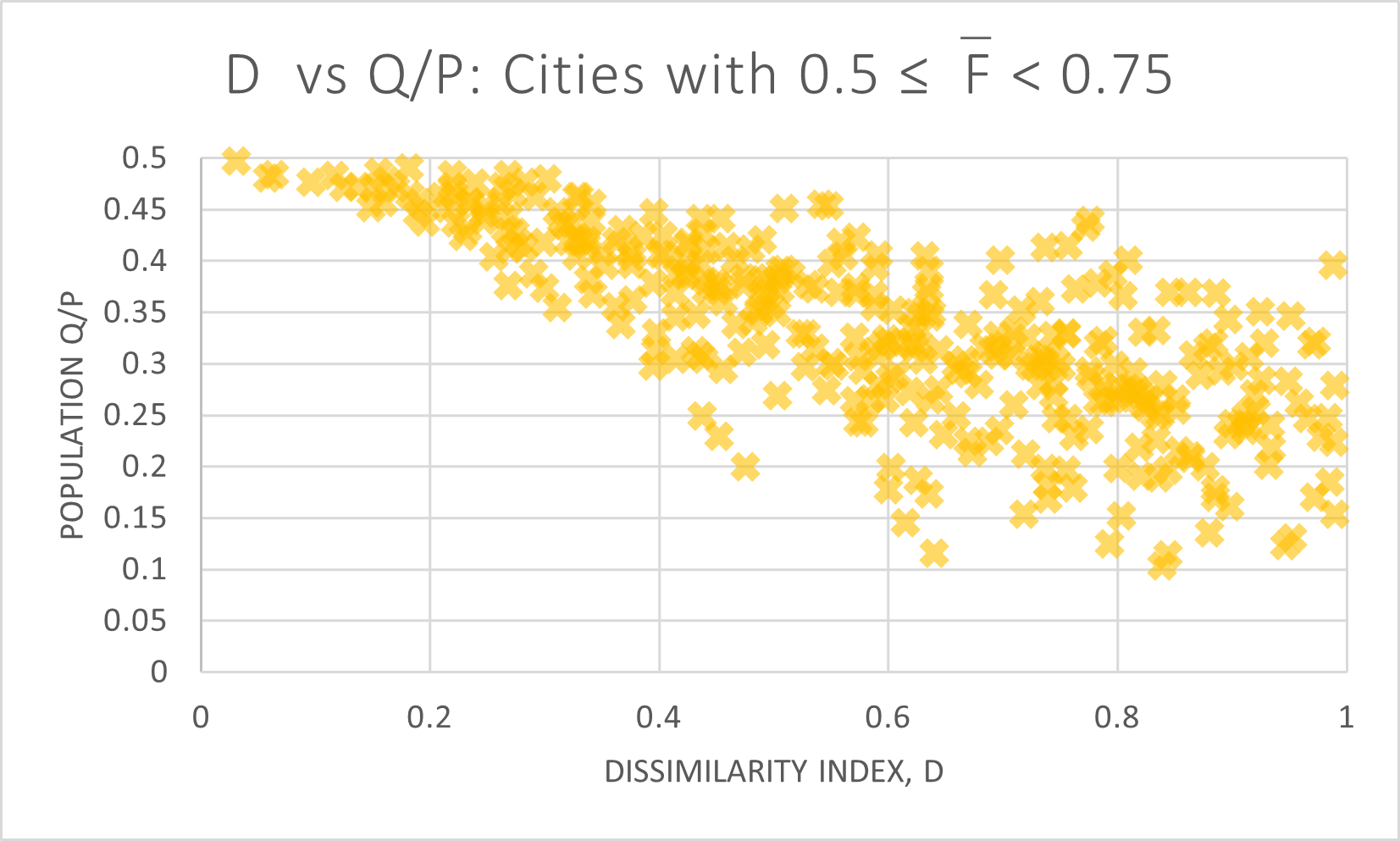}
\includegraphics[scale=.5]{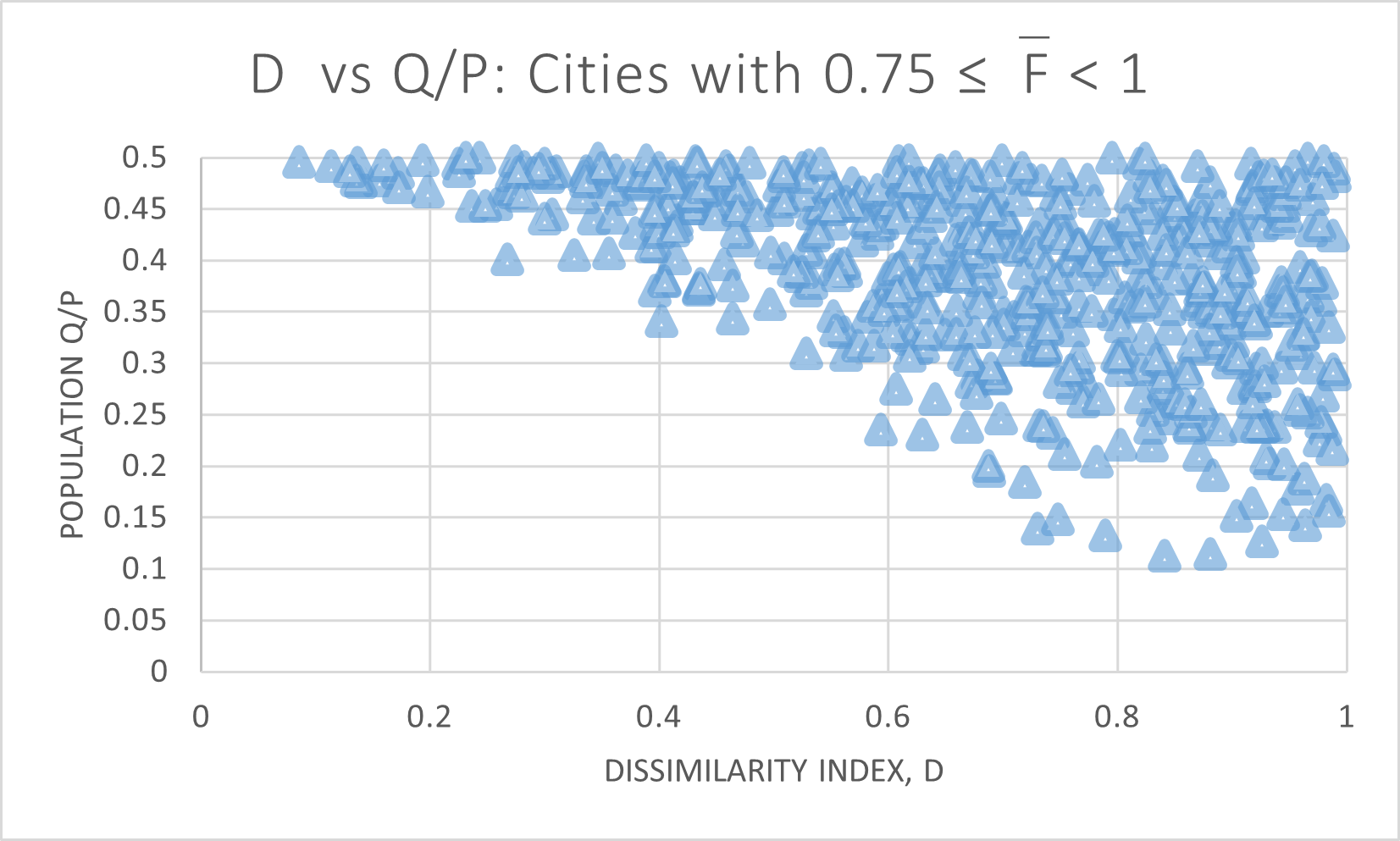}
\includegraphics[scale=.5]{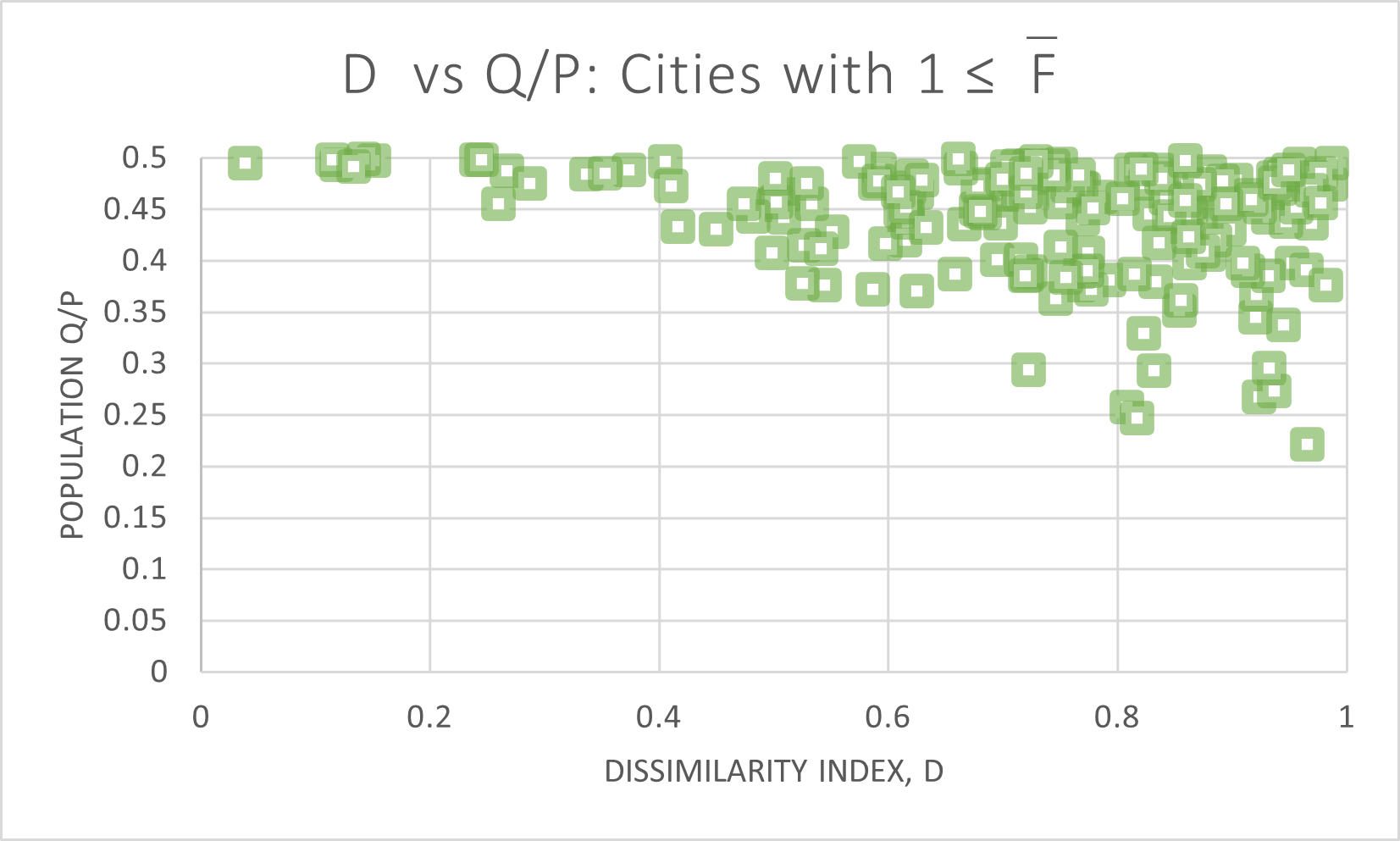}

\caption{Six scatter-plot of Grid Cities by population $Q/P$ and dissimilarity index $D$, one for each of the six $\bar F$ histogram bins.} 
\label{All 6 Scatter-plot Grid Cities by histogram bin}
\end{figure}

For added clarity, we break out each bin individually, in Figure \ref{All 6 Scatter-plot Grid Cities by histogram bin}.

Clearly both $Q/P$ and $D$ contribute towards the availability of a fair district plan. The grid cities with both a large minority population and a fairly high degree of residential segregation have the largest number of proportional district plans available.  However, the vast majority of cities do not have any remotely proportional district plan available.

\subsubsection{Minority percent $Q$ by $\bar F$}

For a fixed segregation index $D$, how does the proportion of $Q/P$ affect $F$? In other words, if segregation patterns are stable but there is an influx or loss by particular groups, how does that affect $F$?

\begin{figure}[h]
\centering
\includegraphics[scale= 0.6]{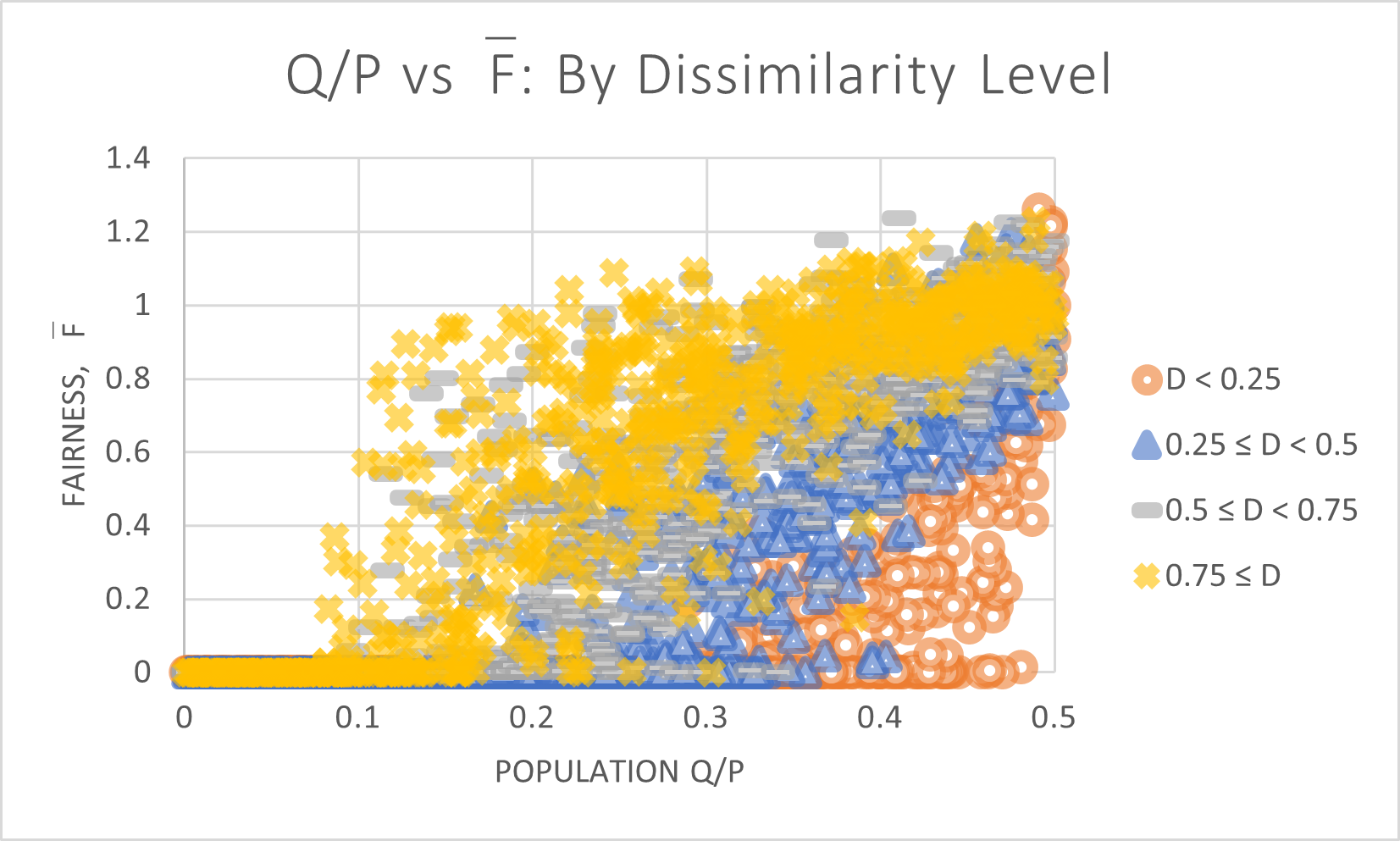}

\caption{Scatter-plot of Grid Cities by population $Q/P$ and fairness index $F$, with colors showing levels of dissimilarity index $D$} 
\label{Scatter-plot Grid Cities by D}
\end{figure}

\begin{figure}[h]
\centering
\includegraphics[scale= 0.5]{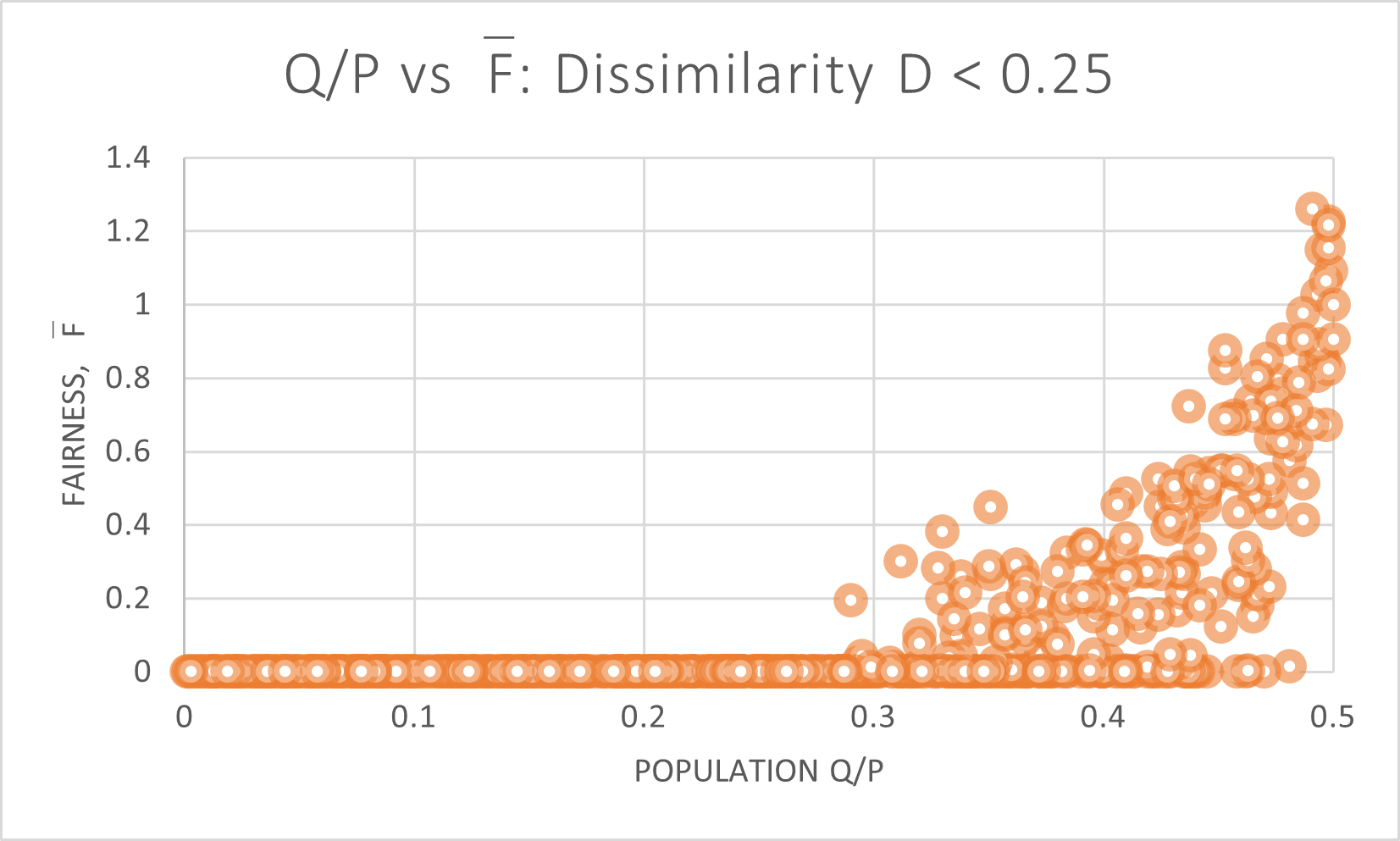}
\includegraphics[scale= 0.5]{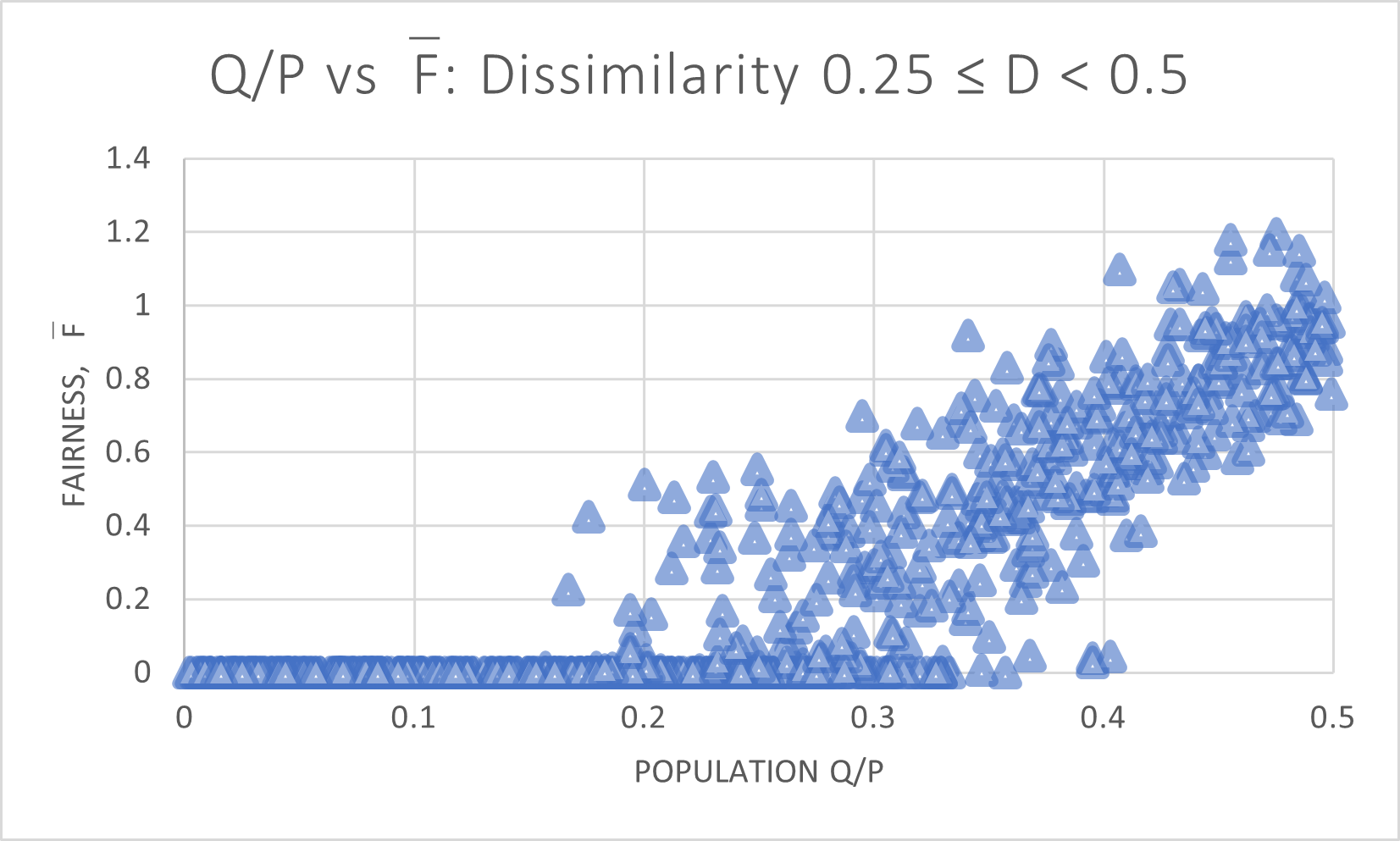}
\includegraphics[scale= 0.5]{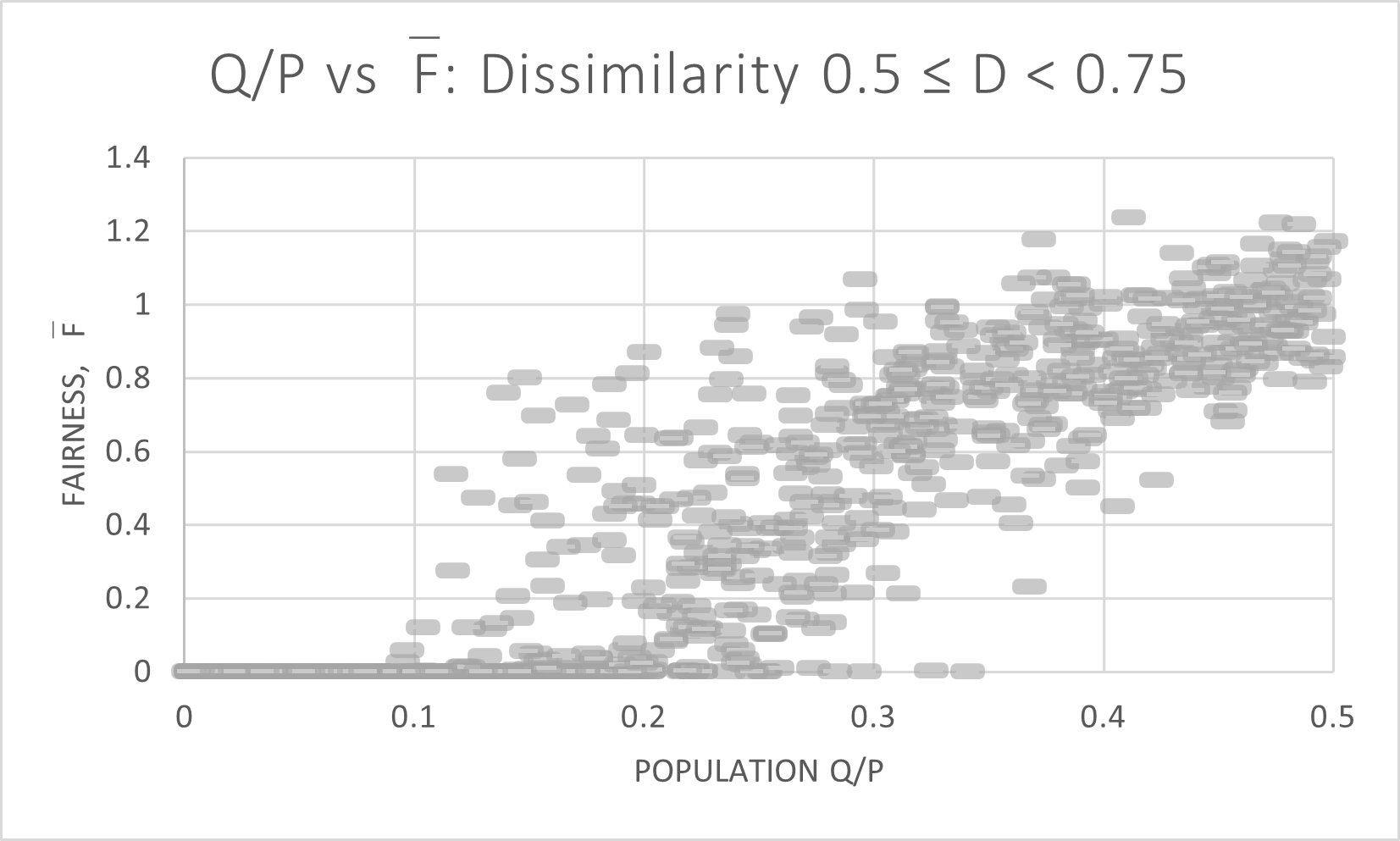}
\includegraphics[scale= 0.5]{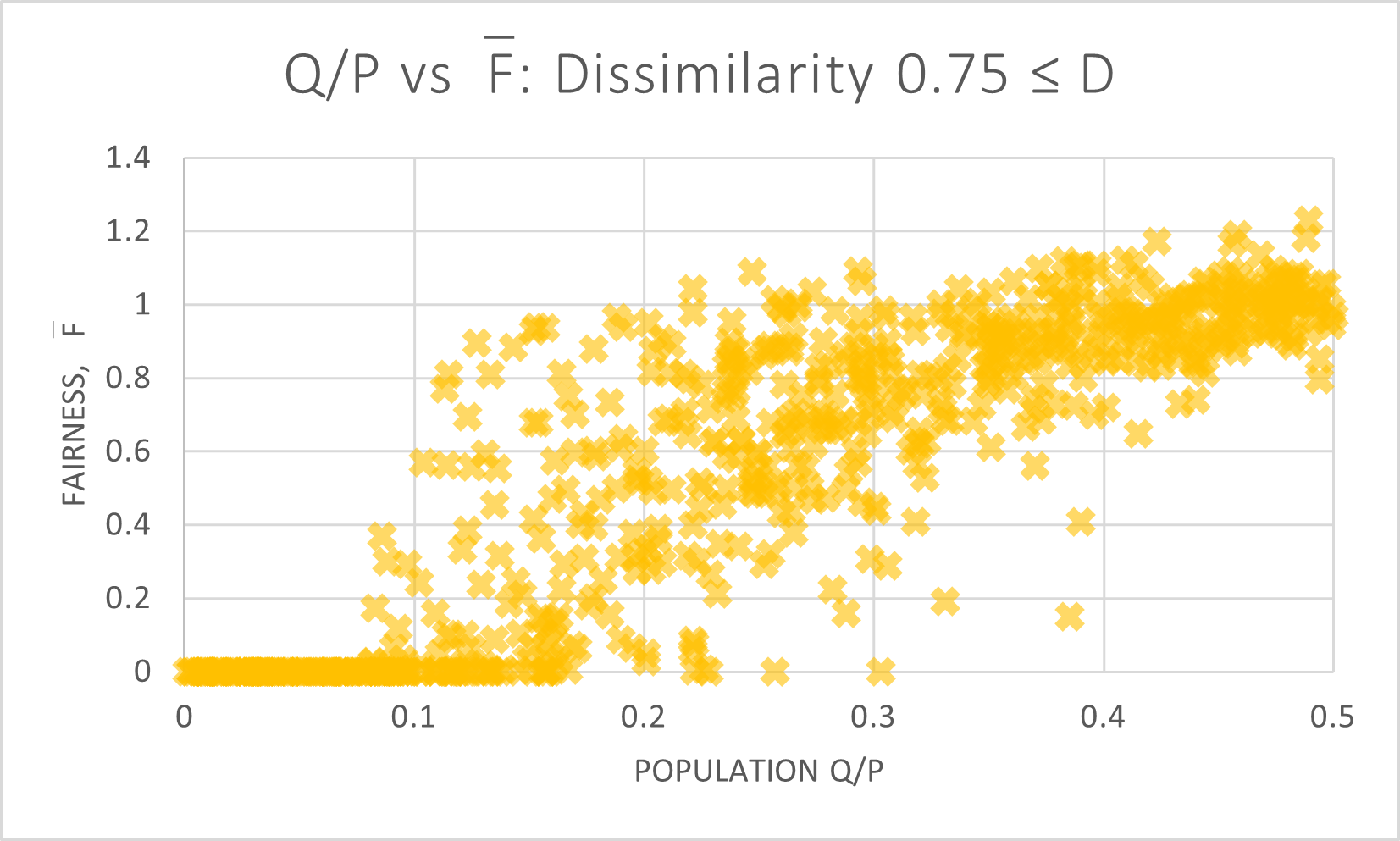}

\caption{Four individual scatter-plots of Grid Cities by population $Q/P$ and fairness index $F$, each restricted to a quartile of the possible range of dissimilarity index $D$} 
\label{Four individ Scatter-plot Grid Cities by D}
\end{figure}

Figure \ref{Scatter-plot Grid Cities by D} shows the distribution of all 3000 cities. At every segregation level, an increase in $Q/P$ correlates strongly with an increase in the fairness index, $\bar F$.
Each level of $D$ is broken out individually in Figure \ref{Four individ Scatter-plot Grid Cities by D}. Regardless of segregation pattern, cities with a smaller minority population will face difficulty finding proportional district plans.

Grid cities with diffuse populations of $Q$ ($D < 0.25$) face dismal prospects for finding any available minority-majority district until they comprise at least $0.4$ of the population.  Highly segregated grid cities ($D \geq 0.75$) have many options for proportional district plans, although it drops off as the minority population decreases below $0.1$.  Between $0.05 < Q/P \leq 0.1$, these communities rarely have any available minority-majority districts, despite still being slightly above the threshhold population needed ($Q/P=0.05$, for 10 districts).

\section{Context within American cities in 2020} \label{context}

In 2020, the dissimilarity index between Black and White populations ranged from 16.01 to 80.04 in 88 American cities with populations over 250,000, with an average $D=45.52$ and standard deviation 14.62.  Between Hispanic and White populations, the range was $10.55 \leq D \leq 63.42$, with an average $D$ of 39.75, and standard deviation 11.37. \cite{LS21} 

\begin{figure} [h]
\centering
\includegraphics[scale=0.6]{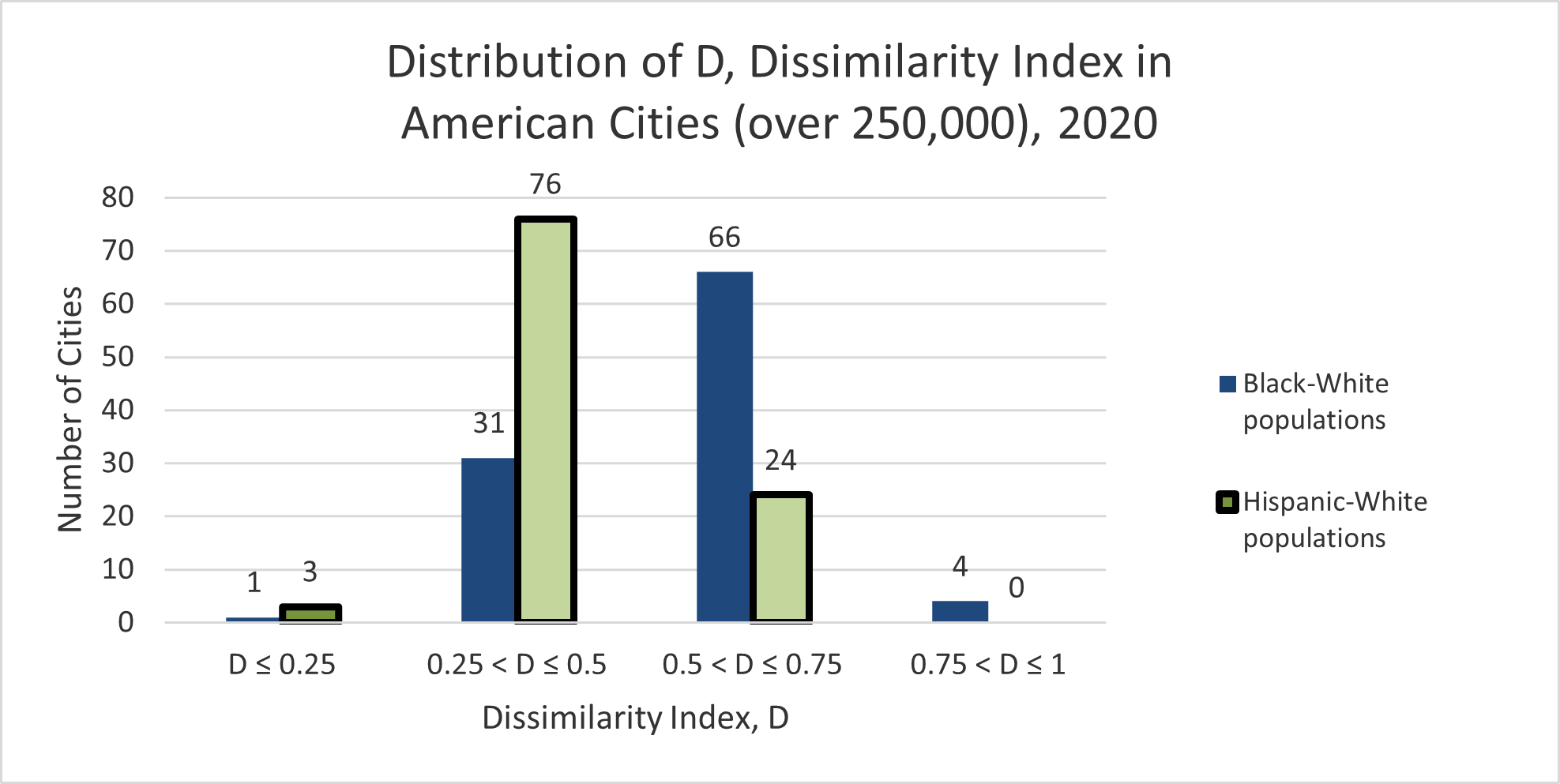}

\caption{Distribution of cities over 250,000 in 2020, by dissimilarity index. Dissimilarity index is shown twice per city, once with respect to Black and White populations, and the other with respect to Hispanic and White populations.} 
\label{Histogram Cities in 2020 over 250K}
\end{figure}

The distribution of $D$ across all cities over 250,000 ($n=88$) are shown in Figure \ref{Histogram Cities in 2020 over 250K}.
Of the cities modeled in this paper, actual values of $D$ in 2010 are:
\begin{itemize}
    \item Albuquerque: 36.4, with respect to Hispanic and White populations
    \item Charlotte: 53.8, with respect to Black and White populations
    \item Pittsburgh: 65.8, with respect to Black and White populations
    \item Minneapolis: 52.9, with respect to Black and White populations
\end{itemize}
\cite{UM-ISR}

\begin{figure} [h]
\centering
\includegraphics[scale=.6]{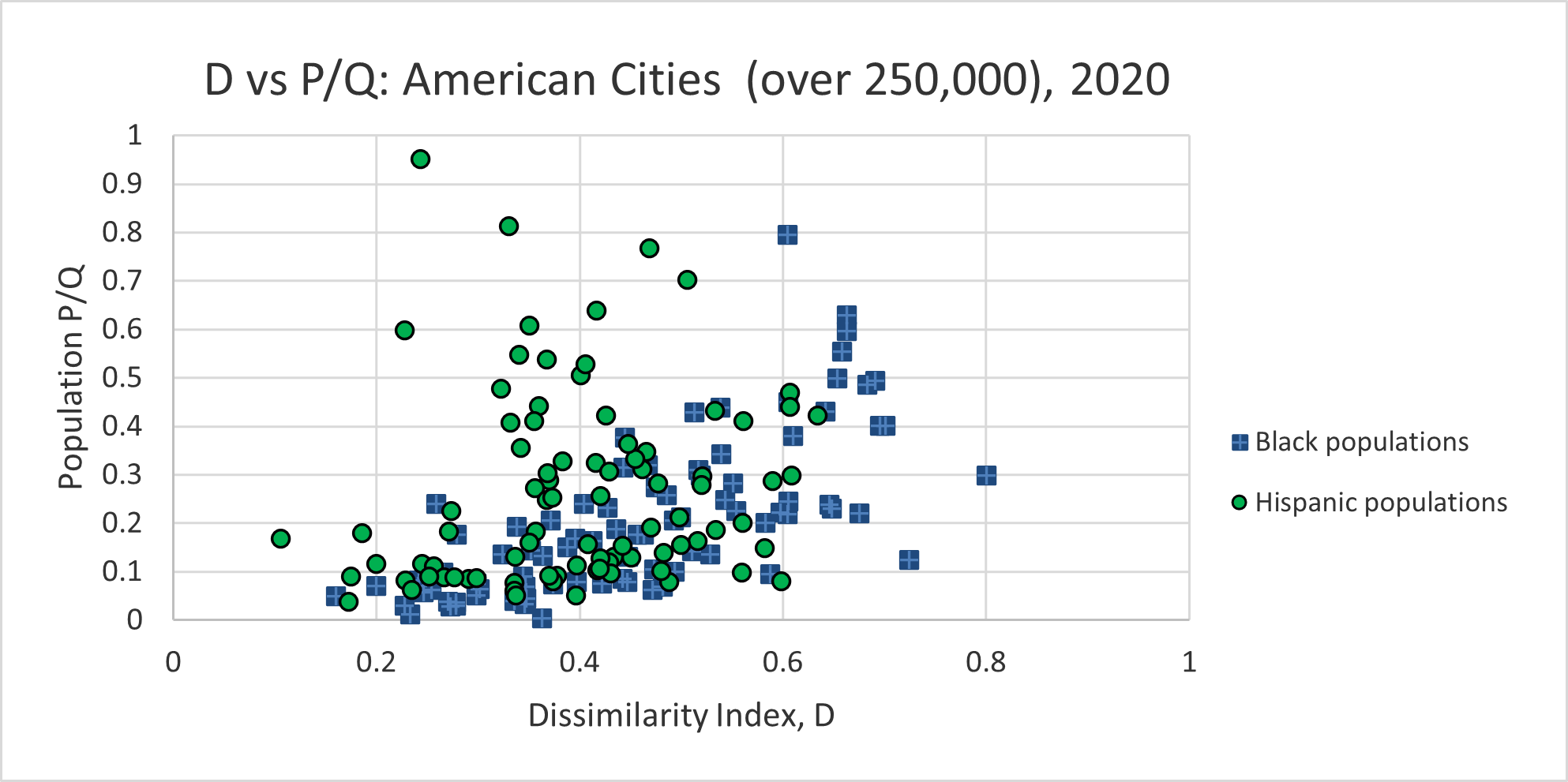}

\caption{Scatter-plot of population $P/Q$ by dissimilarity index $D$, of the cities in 2020 with populations over 250,000. Dissimilarity index is shown twice per city, once with respect to Black and White populations, and the other with respect to Hispanic and White populations. Note that $Q/P$ runs from 0 to 1, since there are cities where the historically vulnerable population is more than 0.50, in contrast to the grid cities.} 
\label{Scatter-plot Actual Cities in 2020 over 250K}
\end{figure}

The full distribution by racial segregation $D$ and percent minority is shown in Figure \ref{Scatter-plot Actual Cities in 2020 over 250K}.

Note that $Q/P$ runs from 0 to 1 here, since there are cities where the historically vulnerable population is more than 0.50. In our grid cities, we designated the smaller population to be $Q$.

\begin{figure} [h]
\centering
\includegraphics[scale=.45]{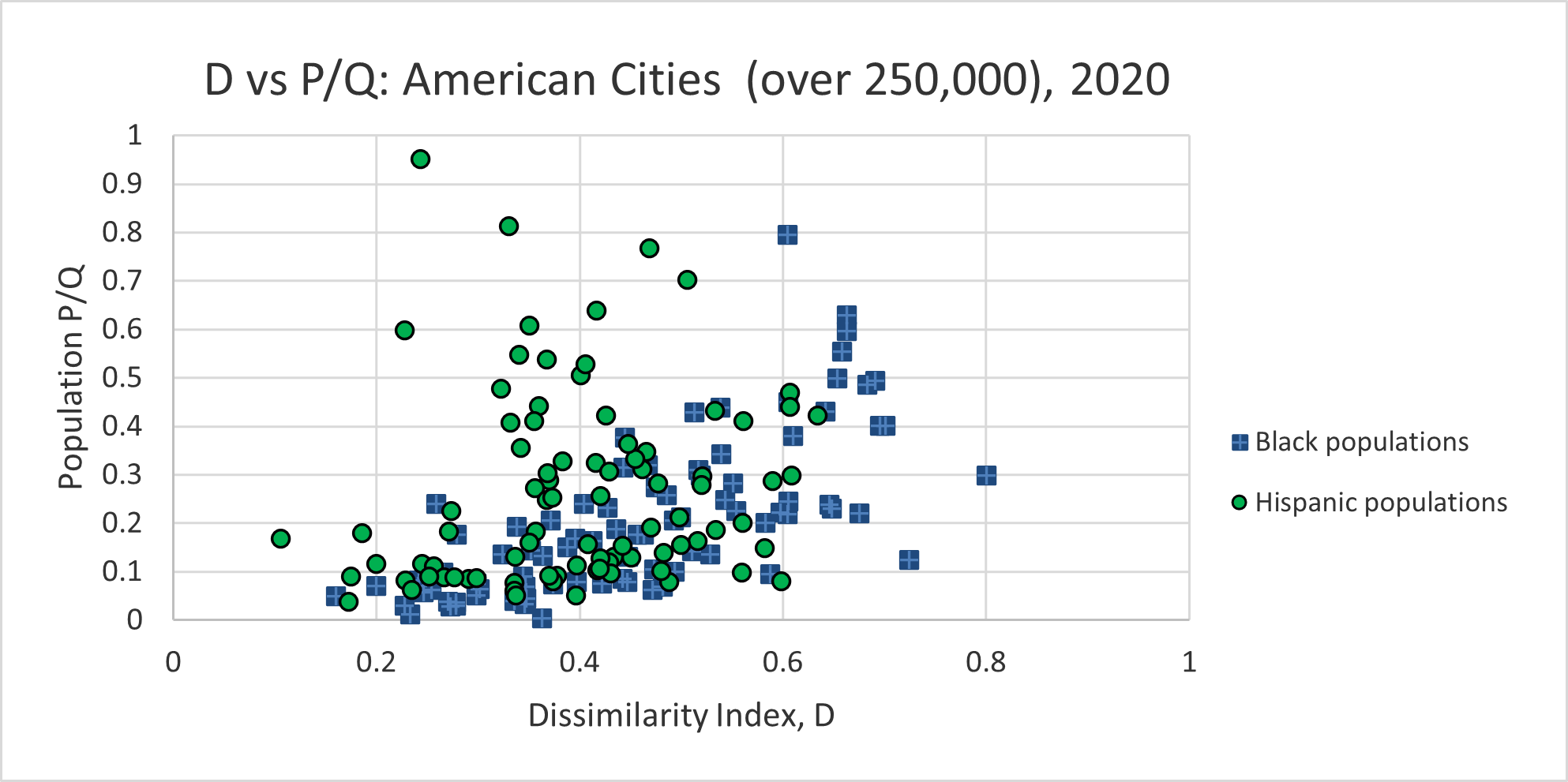}
\includegraphics[scale=0.45]{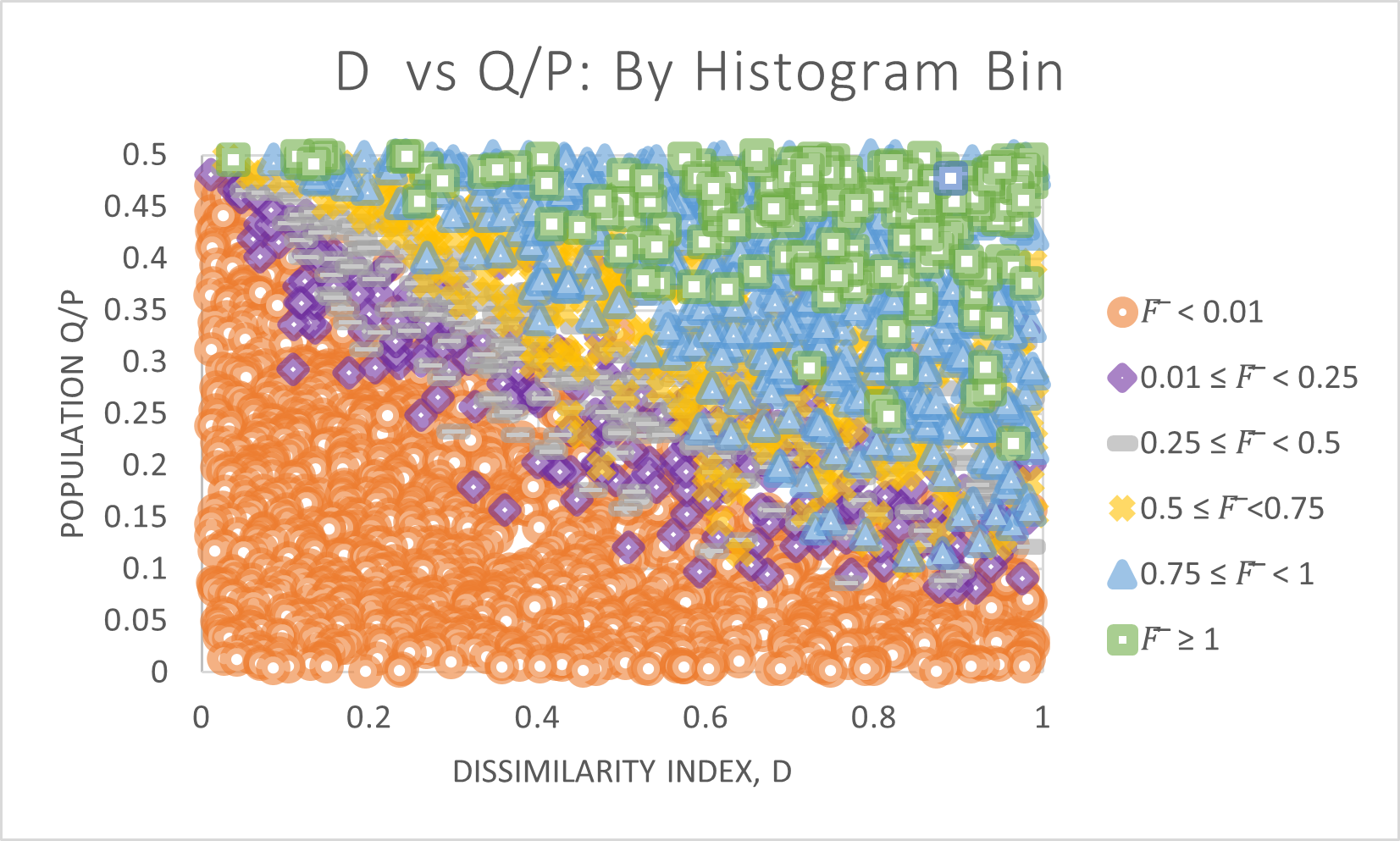}

\caption{Comparison of two scatter-plots of cities, both with respect to population $P/Q$ by dissimilarity index $D$. The first is of American cities over 250,000, and the other is the of the 3000 grid cities. Note that vertical axes are not equivalent, because there exist American cities with $P/Q > 0.5$.} 
\label{Comparison  of Actual cities with Grid Cities}
\end{figure}

If our grid cities are predictive of reality, we would extrapolate the portion of our grid cities that represent the demographics of the actual cities. In Figure \ref{Comparison  of Actual cities with Grid Cities}, the scatter-plot showing the distribution of cities in 2020 is compared with the scatter-plot showing the grid cities.  

(Note that the vertical axes are not equivalent, per the note above.)

Looking at the region occupied by the densest cluster of 2020 American cities, the corresponding grid cities have very low fairnessindices. However, the American cities with higher $Q/P$ and $D$ could be modeled by cities with $\bar F$ much closer to 1.

\section{Conclusion}

We conclude that when the goal is proportional representation, single-member districts are usually a poor mechanism to achieve it, unless populations are very close to 50-50. Other voting systems, such as Ranked Choice Voting or Single Transferable Vote, appear to fare better at this goal. 

We also note that there is occasionally a claim that Democrats self-pack into cities, and thus are underrepresented in district plans because there is no proportional district plan available. \cite{JR13} We are unable to find modeling evidence that suggests that higher concentrations of Democrats would make it more difficult to draw a proportional district plan.

With the release of 2020 Census data, much new analyses can be carried out.  It is likely that soon, most major metropolitan areas will have ensemble analyses carried out to establish a baseline for the availability of a fair district plan.  The ways in which this intersects with racial segregation will be difficult to tease apart, however.  Most major regions have more than two populations which are large enough to constitute a majority of at least one district (depending on the relative size of the district), whereas our model cities were deliberately sought to enable the simplifying assumption of only two sufficiently large racial groups.



\bibliographystyle{plain}
\bibliography{sn-bibliography}

\begin{thebibliography}{10}

\bibitem{RC19}
Rucho et al. v. {C}ommon {C}ause et al., {A}ppeal from the {U}nited {S}tates
  {D}istrict {C}ourt for the {M}iddlea {D}istrict of {N}orth {C}arolina, note =
  {No. 18–422. Argued March 26, 2019—Decided June 27, 2019*}, howpublished
  = {\url {https://www.supremecourt.gov/opinions/18pdf/18-422_9ol1.pdf}},.

\bibitem{Bpedia}
Ballotpedia.
\newblock Thornburg v. {G}ingles.
\newblock \url{https://ballotpedia.org/Thornburg\_v.\_Gingles}.
\newblock Accessed: 2021-12-28.

\bibitem{USCB1}
United States~Census Bureau.
\newblock United {S}tates {C}ensus {B}ureau, {H}ousing {P}atterns, {A}ppendix
  {B}: {M}easures of {R}esidential {S}egregation.
\newblock
  \url{https://www.census.gov/topics/housing/housing-patterns/guidance/appendix-b.html}.
\newblock Accessed: 2021-06-14.

\bibitem{USCBSF}
United States~Census Bureau.
\newblock United {S}tates {C}ensus {B}ureau, census block shapefiles for
  minnesota, new mexico, north carolina, and pensylvania, 2011.
\newblock Retrieved from
  \url{https://www.census.gov/cgi-bin/geo/shapefiles/index.php}.

\bibitem{USCBP4}
United States~Census Bureau.
\newblock United {S}tates {C}ensus {B}ureau, hispanic or latino origin (p4),
  census blocks for minnesota, new mexico, north carolina, and pensylvania,
  2010 decennial census, summary file 1, 2011.
\newblock Retrieved from \url{https://data.census.gov/cedsci/}.

\bibitem{USCBP8}
United States~Census Bureau.
\newblock United {S}tates {C}ensus {B}ureau, race data (p8), census blocks for
  minnesota, new mexico, north carolina, and pensylvania, 2010 decennial
  census, summary file 1, 2011.
\newblock Retrieved from \url{https://data.census.gov/cedsci/}.

\bibitem{JR13}
Jowei Chen and Jonathan Rodden.
\newblock Unintentional gerrymandering: Political geography and electoral bias
  in legislatures.
\newblock {\em Quarterly Journal of Political Science}, 8(3):239--269, June
  2013.

\bibitem{ABQS}
New~Mexico City~of Albuquerque.
\newblock City of albuquerque, city council districts shapefile.
\newblock Retrieved from
  \url{https://www.cabq.gov/gis/geographic-information-systems-data} in 2019.

\bibitem{CHARS}
North~Carolina City~of Charlotte.
\newblock City of charlotte, city council districts shapefile.
\newblock Retrieved from
  \url{https://data.charlottenc.gov/maps/charlotte::council-districts/about} in
  2019.

\bibitem{DDS19}
Daryl DeFord, Moon Duchin, and Justin Solomon.
\newblock Recombination: A family of markov chains for redistricting, 2019.

\bibitem{duchin2018gerrymandering}
Moon Duchin.
\newblock Gerrymandering metrics: How to measure? what's the baseline?
\newblock {\em arXiv: Physics and Society}, 2018.

\bibitem{DGHKNW19}
Moon Duchin, Taissa Gladkova, Eugene Henninger-Voss, Ben Klingensmith, Heather
  Newman, and Hannah Wheelen.
\newblock Locating the representational baseline: {R}epublicans in
  {M}assachusetts.
\newblock {\em Election Law Journal: Rules, Politics, and Policy},
  18-4:388--401, Dec 2019.

\bibitem{DD55}
Otis~Dudley Duncan and Beverly Duncan.
\newblock A methodological analysis of segregation indexes.
\newblock {\em American Sociological Review}, 20(2):210--217, 1955.

\bibitem{GB34}
C.~E. Gehlke and Katherine Biehl.
\newblock Certain effects of grouping upon the size of the correlation
  coefficient in census tract material.
\newblock {\em Journal of the American Statistical Association},
  29:185A:169--170, 1934.

\bibitem{GerryChain}
Metric Geometry and Gerrymandering Group.
\newblock Gerry{C}hain.
\newblock \url{https://github.com/mggg/GerryChain}.
\newblock Accessed: 2021-12-28.

\bibitem{MGGG2}
Metric Geometry and Gerrymandering Group.
\newblock The {K}nown {S}izes of {G}rid {M}etagraphs.
\newblock \url{https://mggg.org/table}.
\newblock Accessed: 2021-12-28.

\bibitem{MINNS}
Minnesota Hennepin~County.
\newblock Hennepin county, ward districts shapefile.
\newblock Retrieved from \url{https://gis-hennepin.hub.arcgis.com/} in 2019.

\bibitem{UM-ISR}
University of~Michigan Institute~of Social~Research.
\newblock New {R}acial {S}egregation {M}easures for {L}arge {M}etropolitan
  {A}reas: {A}nalysis of the 1990-2010 {D}ecennial {C}ensuses.
\newblock \url{https://www.psc.isr.umich.edu/dis/census/segregation2010.html}.
\newblock Accessed: 2021-10-22.

\bibitem{LS21}
John~R. Logan and Brian Stults.
\newblock The persistence of segregation in the metropolis: New findings from
  the 2020 census.
\newblock {\em Diversity and Disparities Project, Brown University}, 2021.

\bibitem{MD88}
Douglas Massey and Nancy Denton.
\newblock The dimensions of residential segregation.
\newblock {\em Social Forces}, 67(2):281–315, 1988.

\bibitem{PITTS}
Pennsylvania Pittsburgh.
\newblock City of pittsburgh, city council districts 2012 shapefile.
\newblock Retrieved from
  \url{https://data.wprdc.org/dataset/city-council-districts-2012} in 2019.

\bibitem{tapp2019measuring}
Kristopher Tapp.
\newblock Measuring political gerrymandering.
\newblock {\em The American Mathematical Monthly}, 126(7):593--609, 2019.

\bibitem{warrington2019comparison}
Gregory~S. Warrington.
\newblock A comparison of partisan-gerrymandering measures.
\newblock {\em Election Law Journal: Rules, Politics, and Policy},
  18(3):262--281, 2019.

\bibitem{W83}
Michael~J White.
\newblock The measurement of spatial segregation.
\newblock {\em American Journal of Sociology}, 88(5):1008–1018, 1983.

\bibitem{YWBM19}
J.~Yao, D.W. Wong, N.~Bailey, and J.~Minton.
\newblock Spatial segregation measures: A methodological review.
\newblock {\em Tijds. voor econ. en Soc. Geog.}, 110:235--250, 2019.

\end{thebibliography}

\section{Appendices}
\appendix

\section{The Algorithm for Creating Synthetic Modeled Cities from Existing Cities}
\label{creating modeled cities}

\subsection{General Description}

The algorithm we use to create synthetic cities from existing cities has several stages.  At the end of the process, we should possess a graph that represents a city where,
\begin{enumerate}
    \item the city-wide population $P$ and city-wide Subgroup $Q$ population are unchanged from the original city, and
    \item the total population within each census block is identical to the original city, even if the Subgroup $Q$ population within each block may be different, and
    \item the dissimilarity index of the synthetic city will be sufficiently close to a chosen target value.
\end{enumerate} 

The intended effect is one where the population of the city is rearranged over the currently existing housing in the city.  This algorithm is used to create a wide variety of subgroup population patterns across the geographies of existing cities.  In particular, a variety of subgroup concentrations and placements is desirable.

\subsection{Creating a Graph of the Existing City}

Before creating a graph of an existing city, a GIS is used to create a shapefile with the necessary geometries and data.  We start by removing all census block polygons that do not intersect a council district polygon.  Next a total population value and one subpopulation value of interest (Black or Hispanic) is joined to the remaining census block polygons.  

To create the city graph, we use the \verb+Graph.from_file()+ method from GerryChain.  This yields a dual graph, where each census block polygon is represented by a vertex and an edges lies between a pair of vertices exactly when the corresponding polygons share some part of their boundary.  All data associated with a polygon in the shapefile is also transferred to its vertex. 

For example, Albuquerque is shown in Figure \ref{Real Abq, part 1},
\begin{figure}[h]
\centering
\includegraphics[scale=.25]{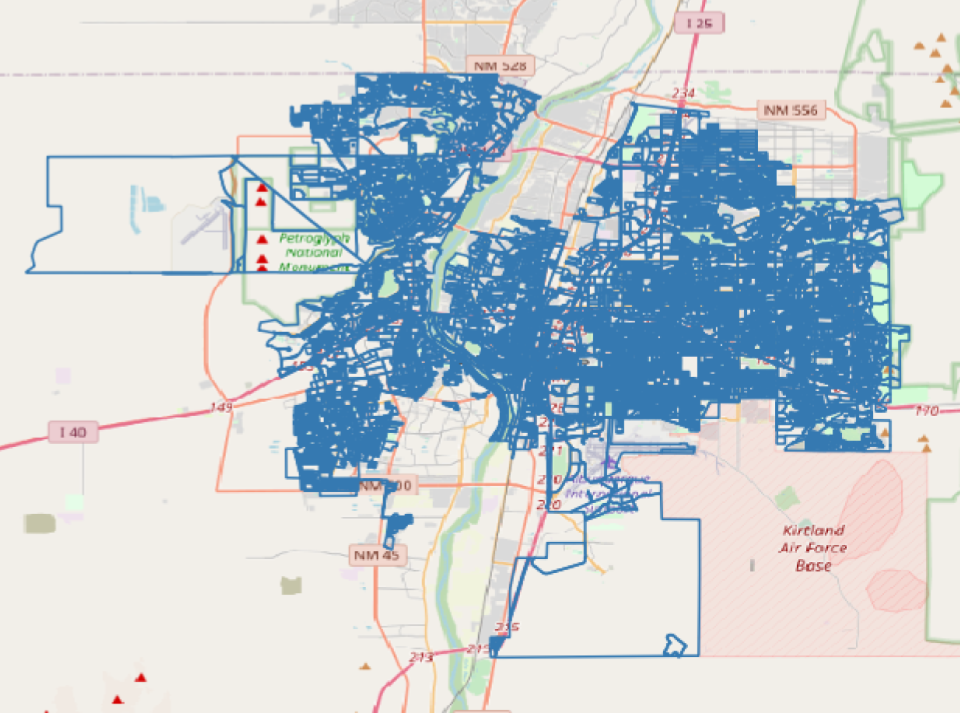}

\caption{Map of Albuquerque, with associated dual graph overlay} 
\label{Real Abq, part 1}
\end{figure}
with corresponding dual graph in Figure \ref{Real Abq, part 2}.
\begin{figure}[h]
\centering
\includegraphics[scale=.25]{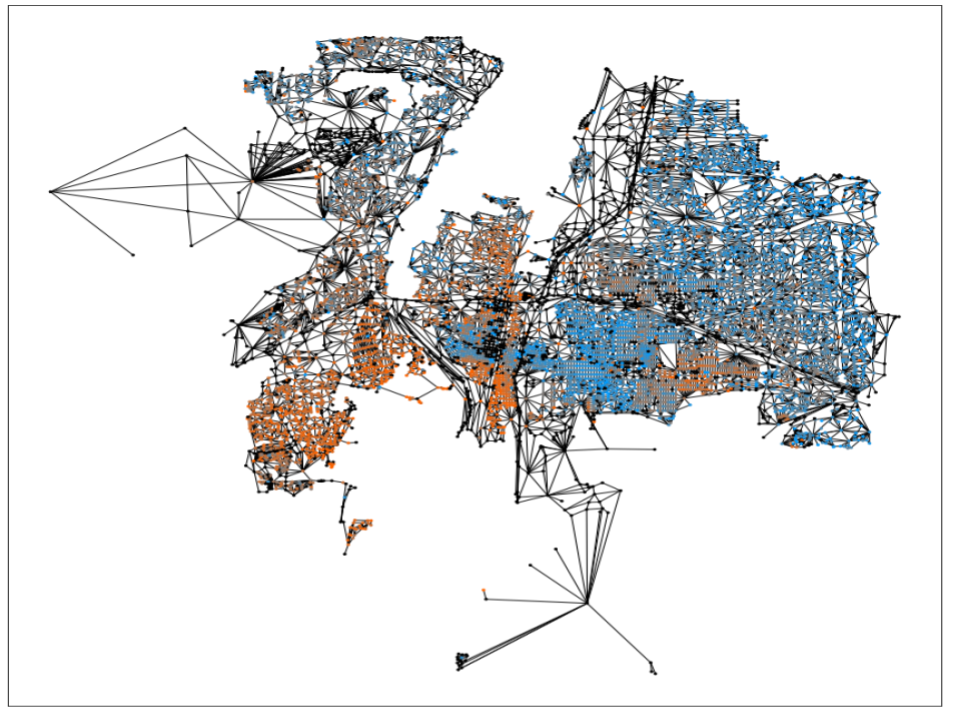}

\caption{Dual Graph of Albuquerque} 
\label{Real Abq, part 2}
\end{figure}
In the latter, the orange and blue represent the distribution of Hispanic individuals (orange) and non-Hispanic individuals (blue), according to the 2010 census.

\subsection{Creating Weights for Population Placement}

After the graph of the city is built, we will assign two weights to each vertex of the city graph: one for Subgroup $Q$ and one for Subgroup $P-Q$.

First, we designate clusters of nodes (separately for Subgroups $Q$ and $P-Q$) on which to place the highest weights.  Between 1 and 10 highest weight clusters are generated for each subgroup across the city.   Highest weights are placed on enough nodes in these clusters to house 80\% of Subgroup $Q$ and 20\% of Subgroup $P-Q$. This is the step in the algorithm which treats $Q$ and $P-Q$ in an asymmetric manner.

Once all of the highest weight clusters are in place, the weights are tapered down across adjacent vertices until reaching a city-wide minimum.  The slope of this taper is determined randomly.  A shallow taper makes that cluster into a broader region for the subpopulation. This may result in a more diffuse population pattern for that subgroup.  A steeper taper makes the cluster into a more tightly focused region. This may result in more concentrated population patterns for that subgroup.

\subsection{Placing Subgroup Populations}

Once Subgroup $Q$ and Subgroup $P-Q$ have weights assigned to each node of the city, the algorithm proceeds to assign subgroup populations randomly to the vertices of the city graph according to these weights.  Subgroups $Q$ and $P-Q$ are offered a list of available vertices, weighted accordingly, to place their members.  The subgroups alternately place groups of up to 20 members on vertices until their total population has been reached.  As vertices reach capacity, they are removed from the list of available vertices.

\subsection{Adjusting the Dissimilarity Index}

The random placements of Subgroups $Q$ and $P-Q$ in the previous step yield a city graph with a dissimilarity index that is likely different from our desired target index.  We now use a list of graph vertices sorted by Subgroup $Q$ population percentage to adjust the index toward the target.  If dissimilarity index must be increased to reach the target, then people from Subgroup $Q$ in vertices with low $Q$-percentage are swapped with people from Subgroup $P-Q$ in vertices with high $Q$-percentage so that the $Q$-percentage ordering of the vertices changes little.  Similarly, to lower the dissimilarity index, people from Subgroup $P-Q$ in vertices with low $Q$-percentage are swapped with people from Subgroup $Q$ in vertices with high $Q$-percentage.  Repeat this process until the graph dissimilarity index is sufficiently close to the target.  Since the $Q$-percentage ordering of the vertices remains largely unchanged, the locations of concentrations of Subgroup $Q$ remain relatively undisturbed. 

After this process is complete, each vertex in the city graph will contain the same number of people as its counterpart in the graph of the original city.  The city-wide populations of Subgroup Q and Subgroup $P-Q$ will remain the same as well. Further, the dissimilarity index of the city will be sufficiently close to the desired target.

\section{The Algorithm for Creating Grid Cities}
\label{creating grid cities}

\subsection{General Description}

Our grid cities are built on $30 \times 30$ grids.  Each vertex of our grid holds 1000 people.  Each person belongs to either Subgroup $Q$ or Subgroup $P-Q$.  

Before any populations are assigned to the grid city, a target Subgroup $Q$ percentage and target dissimilarity index are randomly generated.  We wish to generate a city that meets these targets.

\subsection{Placing Subgroup Populations}

We start populating our $30 \times 30$ grid by distributing people in Subgroup $Q$ uniformly over all of the vertices of the graph.  The remaining space in each vertex is populated by people from Subgroup $P-Q$.  This results in a $30 \times 30$ grid with dissimilarity index $0$.

We then use our algorithm, described in the previous section, to redistribute populations $Q$ and $P-Q$ around the grid in a way that achieves a target dissimilarity index value.

\section{Assembling the Ensemble of District Plans}

\subsection{Number and Type}  In this study 20,003 district plans were used for our grid city and 20,001 were used in each of our model cities of Albuquerque, Charlotte, Pittsburgh, and Minneapolis.  A district plan with $n$ single-member districts is defined by a partition of the census blocks into $n$ equal classes,  such that each class is connected, and the total populations in each class are within 0.2 of $P/n$.

For the grid cities:
\begin{enumerate}
    \item Five partitions were generated ``from scratch" using the algorithm in Appendix \ref{district plans from scratch}.
    \item 19,996 partitions were generated using GerryChain's ``recom'' proposal.\cite{GerryChain}
    \item Two were drawn by hand.
\end{enumerate}

For the four model cities:
\begin{enumerate}
    \item Five partitions were generated ``from scratch" using an algorithm in Appendix \ref{district plans from scratch}.
    \item 19,996 partitions were generated using GerryChain's ``recom'' proposal.
\end{enumerate}

In each city GerryChain was used to generate 4,999 partitions from four different ``from scratch'' partitions, yielding the 20,000 partitions we used for the ensemble analysis.  The remaining partitions (three for the grid city, one for each model city) were used to validate the algorithm runtime for the other 20,000.   

\subsection{Generating Partitions ``from scratch''}The algorithm for creating ``from scratch'' partitions functions as follows: 
\label{district plans from scratch}
\begin{enumerate}
    \item Randomly assign one vertex in the city graph to each district.
    \item Every unassigned vertex adjacent to an assigned vertex should be given to the assigned vertex's district.  Repeat until all vertices are assigned a district.  Note that districts will be now be contiguous, but are unlikely to have equal populations.
    \item Have the district with the largest population give its boundary vertices to adjacent districts.  The populations between the districts will probably be closer to equal, but districts are no longer guaranteed to be contiguous.
    \item Any noncontiguous districts will give away the boundary vertices of its smaller components to adjacent district until the smaller components are gone.  Note that populations may not be sufficient close to equal at this point, but all districts will be contiguous.
    \item Repeat steps 3 and 4 until all districts are contiguous and fall within population constraints.
\end{enumerate}

Note that it is possible for this algorithm to get stuck repeating steps 3 and 4. In our implementation, this ended up being a rare occurrence

\subsection{GerryChain Settings}
\label{gerrychain settings}
Here are the GerryChain settings we used to create the 19,996 partitions for each city in this experiment.  Please refer to GerryChain documentation for an explanation. \cite{GerryChain}

We used GerryChain's MarkovChain command with the following settings:
\begin{itemize}
    \item proposal=recom
    \item constraints=[contiguous, pop\_constraint]
    \item accept=always\_accept
\end{itemize}

Note that pop\_constraint requires the population of each district to differ from the mean by no more than 20\%.

\section{Data Availability Statement}

Most data used for and generated by this project is available to the reader, with the exception of city council district plans that have since been updated by their respective municipalities.

\subsection{External Data Sources}
We used the following sources of data for this project.
\begin{enumerate}
    \item Census block shapefiles downloaded from \url{https://www.census.gov/cgi-bin/geo/shapefiles/index.php}
        
    \item City/county district shapefiles downloaded from Albuquerque, Charlotte, Hennepin County (Minneapolis), and Pittsburgh geodata sites.
    \begin{enumerate}
        \item  Albuquerque City Council Districts (Posted version updated for 2022. Accessed in 2019.):  \url{https://www.cabq.gov/gis/geographic-information-systems-data}
        \item Charlotte City Council Districts (Posted version updated for 2022. Accessed in 2019.): \url{https://data.charlottenc.gov/maps/charlotte::council-districts/about}
        \item Hennepin County Wards (Contains Minneapolis.  Posted version updated for 2022.  Accessed in 2019.):  \url{https://gis-hennepin.hub.arcgis.com/}  
        \item Pittsburgh City Council Districts (2012 map.  Accessed 2019):  \url{https://data.wprdc.org/dataset/city-council-districts-2012}
    \end{enumerate}
    \item Population data CSV files by census block from \url{https://data.census.gov/cedsci/}
\end{enumerate}

\subsection{Research Data}
The following data generated by our research is available at the link \url{https://osf.io/br5tf}.
\begin{enumerate}
    \item Dual graphs of the Albuquerque, Charlotte, Minneapolis, Pittsburgh, and our grid city.  These can be found in the ``Dual Graph'' folder for each city.
    \item District assignments for each dual graph vertex in each of the 20,000 district plans generated for each city's ensemble.  These can be found in the ``Ensemble District Assignments'' folder for each city.
    \item District assignments for each dual graph vertex of the reference district plan used to validate each city's ensemble.  These can be found in the ``Validation Reference District Assignments'' folder for each city.
    \item Population data for each vertex of each generated city, along with dissimilarity index values and mean number of opportunity districts, across 20,000 district plans, for each city.  These can be found in the ``Population Placement and Analysis'' folder.
\end{enumerate} 



\end{document}